\pdfoutput=1
\documentclass[12pt]{article}
\usepackage[margin=1in]{geometry}
\usepackage{amsmath}
\usepackage{titling}
\usepackage{blindtext}
\usepackage{pgfplots}
\usepackage{natbib}
\def\sin{\mathop{\rm sin}\nolimits}
\usepackage{comment}
\usepackage[colorlinks=true,linkcolor=blue,allcolors=blue]{hyperref}
\usepackage{setspace}
\usepackage{authblk}
\begin{document}

\title{Data-constrained Model for Coronal Mass Ejections Using Graduated Cylindrical Shell Method}


\author[1]{T. Singh}
\affil[1]{Department of Space Science, The University of Alabama in Huntsville, AL 35805, USA}

\author[2]{M. S. Yalim}
\affil[2]{Center for Space Plasma and Aeronomic Research, The University of Alabama in Huntsville, AL 35805, USA}

\author[1,2]{N. V. Pogorelov}
\setcounter{Maxaffil}{0}
\renewcommand\Affilfont{\itshape\small}
\date{}    
\begin{titlingpage}
    \maketitle
\begin{abstract}

Coronal Mass Ejections (CMEs) are major drivers of extreme space weather conditions, this being a matter of serious concern for our modern technologically-dependent society. Development of numerical approaches that would simulate CME generation and propagation through the interplanetary space is an important step towards our capability to predict CME arrival times at Earth and their geo-effectiveness. In this paper, we utilize a data-constrained Gibson--Low (GL) flux rope model to generate CMEs. We derive the geometry of the initial GL flux rope using the Graduated Cylindrical Shell (GCS) method. This method uses multiple viewpoints from \textit{STEREO A \& B} Cor1/Cor2, and \textit{SOHO}/LASCO C2/C3 coronagraphs to determine the size and orientation of a CME flux rope as it starts to erupt from the Sun. A flux rope generated in this way is inserted into a quasi-steady global magnetohydrodynamics (MHD) background solar wind flow driven by \textit{SDO}/HMI line-of-sight magnetogram data, and erupts immediately. Numerical results obtained with the Multi-Scale Fluid-Kinetic Simulation Suite (MS-FLUKSS) code are compared with \textit{STEREO} and \textit{SOHO}/LASCO coronagraph observations in particular in terms of the CME speed, acceleration, and magnetic field structure.


\end{abstract}
\end{titlingpage}



\section{Introduction} \label{sec:intro}

Coronal Mass Ejections (CMEs) are the most energetic events in our solar system. They are large structures of plasma confined in a sheared/twisted magnetic 
field being ejected from the low solar corona. Generally, they originate from the  magnetically active regions of the Sun. With ejected mass reaching  $10^{12}$ kg and speeds up to 3000 km/s, they carry a huge amount of kinetic and magnetic energy~\citep{Chen11}. A CME directed towards Earth can cause extreme space weather conditions that affect space-borne and ground-based technological systems. Therefore, predicting the CME eruption, its arrival time at Earth, and possible impact on it are of great importance to our technologically-advanced society. Many past and present observatories and instruments (both space-borne and ground-based) have helped us understand the Sun--Earth connection. A number of CME arrival time models have been proposed over the years. They include empirical models \citep[e.g.][]{Vandas96,Brueckner98,Gopalswamy05,Wang02,Manoharan04}, drag based models used to predict CME arrival times~\citep{Vrsnak01,VG02}, and such physics based-models such as, e.g., Shock Time of Arrival (STOA), STOA-2~\citep{MDS02}.

Substantial success has been achieved in numerical modeling of CMEs \citep[e.g.][]{Amari11,ACA14,ADK99,Aulanier10,Roussev12,Jiang16,Forbes06, ML94,Moore01,Schmieder15,TK05,KT06,GL98,TD99,
Titov14,LM95,Chane05,Chen11,JPH06,Jin17a,Lugaz17,TKT04,FG07,FP95,LF00,Hu01} and their propagation into the inner heliosphere \citep[e.g.][]{Detman11,vanderHolst14,Feng11,FMX15,Hayashi13,Intriligator12,LLL14,LLA14,Linker16,LLM09,Lionello16,
LR11,Manchester06,Merkin16,OP99a,OP09,Oran15,RML03,RLA15,
Riley15,RR13,Roussev03b,Sokolov13,UG06,Usmanov11,Wang11,Wu09}. 

Previous CME models such as the blob model and over-pressured spherical plasmoid~\citep[e.g.][]{Chane05,OP99a} do not take into consideration the magnetic field inside a CME. These approaches do not give us a complete picture of CME propagation because the conversion from magnetic to kinetic energy is an integral part of this phenomenon. Processes like CME-CME collisions in the interplanetary space rely heavily on the CME magnetic field. Thus, the above models fail to simulate the full complexity of CME events~\citep{Shen17}. The magnetic field produced by a CME is one 
of the critical parameters determining its geoeffectiveness, i.e., the ability to disturb Earth's magnetosphere and upper atmosphere. CMEs with a negative  z-component of the magnetic field vector,  $B_z$, have been observed to be more geoeffective due to coupling with the positive $B_z$ of Earth's magnetosphere, where the z-axis is perpendicular to the solar ecliptic plane~\citep{Lockwood16}. Thus, CME models, that ignore such magnetic structure can hardly be used to predict their geoeffectiveness.

In this paper, we use a Gibson--Low (GL) type flux rope model~\citep{GL98} to simulate a CME. Similar models have previously been applied by, e.g.,~ \citet[][]{ Manchester04a,Manchester04b,Manchester06,Manchester14a,MVL14,Jin16,Jin17a,Jin17b,Kataoka09,Lugaz05,Lugaz07,PP17,Pomoell17,SK16}. 
\citet{Jin17b} describe a data-constrained CME model to find the GL flux rope parameters from observations. They use the size of neutral line in the source active region to find the GL size parameters. The GL magnetic field strength is found indirectly from a parametric study.

In the present paper, we acquire the GL flux rope size parameters directly from coronagraph observations of a CME by using the Graduated Cylindrical Shell (GCS) method~\citep{THV06}. Afterwards, a parametric study is performed to compute the magnetic field strength of flux rope indirectly. This method is described in detail in section~\ref{models}.

This approach is complementary to the CME model of~\citet{Jin17b}. It allows us to determine the initial flux rope geometry more accurately because we do not impose excessive energy in the initial flux rope configuration thereby avoiding its excessive heating and acceleration. Moreover, our method of determining the GL flux rope parameters from the observational data can be automatized by a user friendly GUI similar to the Eruptive Event Generator (Gibson and Low)(EEGGL)~\citep{Borovikov17} in the Community Coordinated Modeling Center (CCMC). While complex CME models involving an energy buildup before eruption~\citep[e.g.][]{TD99,ACA14} exist, our model implements a rather simple, but data driven, eruption mechanism triggered by the force imbalance between the initial flux rope and the surrounding background solar wind as soon as the flux rope is inserted.  As compared with a number of CME initiation models described in the reviews of ~\citet{Chen11} and ~\citet{Aulanier13}, especially taking into account existing limitations on data-driven models, our approach is computationally more efficient and provides a practical 
alternative for operational space weather forecasting.

We have implemented this CME model as a module in the Multi-Scale Fluid-Kinetic Simulation Suite (MS-FLUKSS)~\citep{Pogorelov14}-- a suite of adaptive mesh refinement (AMR) codes designed to solve the coupled system of magnetohydrodynamics (MHD), gas dynamics Euler, and kinetic Boltzmann equations~\citep{Borovikov09,Borovikov13,Pogorelov09,Pogorelov13}. MS-FLUKSS is built upon the Chombo AMR framework~\citep{Colella07}. It also has modules that treat pickup ions either kinetically or as a separate fluid, and turbulence models applicable beyond the Alfv\'enic surface~\citep{Gamayunov12,Kryukov12,Adhikari15}.

Previously, we have studied a number of CME events generated by the blob model~\citep{Chane05} in the inner heliosphere using MS-FLUKSS~\citep{Pogorelov17}. Our present CME model employs a newly developed data-driven MHD global solar corona model~\citep{YPL17}.  

The structure of the paper is as follows. In section~\ref{models}, we present an overview of our global solar corona and CME models. In section~\ref{res}, we present our numerical results. Finally, in section~\ref{conc}, we draw some conclusions pertinent to our simulations.

\section{Models} \label{models}

\subsection{Global Solar Corona Model} \label{swmodel}

There have been a few attempts to obtain flux ropes in solar corona suitable for CME generation. Worth mentioning, in particular, is the magnetofrictional method~\citep{Cheung15,Fisher15}. \citet{Jiang16} reported a CME born at an active region on the solar surface on basis of the MHD conservation laws with appropriate plasma heating mechanism similar to the one used in this paper. We are also pursuing similar approaches. However, they have been applied so far only to localized active regions. The difficulty is to ensure that such structures create CMEs only when they are observed. A simplified alternative is to insert a flux rope defined by analytical solutions into a previously obtained, background solar wind flow propagating towards Earth. This imposes critical restrictions onto any background model, since otherwise even a perfect CME model may lead to inaccurate results. On the other hand, oversimplified models of CME propagation may show excellent agreement with observed CME shock arrival time at Earth when they propagate through the background solar wind which disagrees with in situ observations during quiet-Sun periods. 

For this reason, we have developed a new, data-driven global MHD model of solar corona and inner heliosphere~\citep{YPL17}, which is based on vector magnetograms and therefore makes it possible to implement mathematically consistent, characteristics-based boundary conditions. Since we are solving the system of hyperbolic MHD equations, the boundary conditions in lower corona should be specified according to the theory of characteristics. 

Consider for simplicity a 1D system of conservation laws
\begin{equation}
\frac{\partial \mathbf{U}}{\partial t} +\frac{\partial\mathbf{F}}{\partial x}=0,
\end{equation}   
where $\mathbf{U}$ and $\mathbf{F}$ are the vectors of conservative variables and corresponding fluxes, respectively. 

This system can be rewritten in a quasi-linear form as

\begin{equation}
\frac{\partial \mathbf{U}}{\partial t} + A\frac{\partial\mathbf{U}}{\partial x}=0, \quad
A=\frac{\partial\mathbf{F}}{\partial \mathbf{U}}.
\end{equation}

Since the MHD system is hyperbolic, the Jacobian matrix $A$ has only real eigenvalues, $\lambda_i,
i=1,\ldots,8$. Moreover, there exists a non-degenerate, complete set of left and right eigenvectors for this matrix, i.e.,
\begin{equation}
A\Omega_\mathrm{R}=\Lambda \Omega_\mathrm{R}, \quad
\Omega_\mathrm{L} A= \Lambda \Omega_\mathrm{L},
\end{equation}
where $\Omega_\mathrm{R}$ and  $\Omega_\mathrm{L}$ are the matrices formed by the right and left eigenvectors of A, used as columns and rows, respectively. In addition, $\Lambda$ is a diagonal matrix formed of eigenvalues of $A$.\\

From the above it follows that
\begin{equation}
A=\Omega_\mathrm{R}\Lambda\Omega_\mathrm{L}.
\end{equation}

On introducing the vector $\mathbf{w}$ such that $d\mathbf{w}=\Omega_\mathrm{L} d\mathbf{U}$, we obtain
\begin{equation}
\frac{\partial\mathbf{w}}{\partial t}+ \Lambda \frac{\partial\mathbf{w}}{\partial x}=0,
\end{equation}
or
\begin{equation}
\frac{\partial w_i}{\partial t} +\lambda_i \frac{\partial w_i}{\partial x}=0, \quad i=1,\ldots,8,
\label{char}
\end{equation} 
where $\mathbf{w}=[w_1,w_2,\ldots, w_8]^\mathrm{T}$.

We implicitly assumed here that the $x$-axis is perpendicular to a chosen boundary of the computational regions. E.g., it can coincide with the radial direction on a spherical inner boundary placed into the lower corona. 

It is clear from Eq.~\ref{char} that the propagation of each $w_i$, which are called the characteristic variables, is described by an independent transport equation. Each of these equations are convection equations describing the propagation of $w_i$ with the speed $\lambda_i$ along the characteristic path $dx/dt=\lambda_i$. 

Thus,  physical boundary conditions should be specified only for characteristic variables that enter the computational region. For the entrance boundaries, this corresponds to $\lambda_i>0$. For the system of ideal MHD equations, we have eight eigenvalues:
\begin{equation}
\lambda_{1,2}=u, \ \lambda_{3,4}=u\pm c_\mathrm{s},\  \lambda_{5,6}=u\pm c_\mathrm{A},
\ \lambda_{7,8}=u\pm c_\mathrm{f},
\end{equation} 
where $c_\mathrm{s}$, $c_\mathrm{A}$, and $c_\mathrm{s}$ are the slow magnetosonic, Alfven, and fast magnetosonic speeds, respectively.

Thus the number of boundary conditions is not arbitrary and depends on the number of positive eigenvalues. Such boundary conditions are called physical. The rest of boundary conditions are mathematical. Clearly, only certain components of the vector of characteristic variables should be specified as physical. Unfortunately, there are no analytic expressions for $w_i$ in MHD. In addition, one would prefer to specify measurable quantities as physical boundary conditions. For this to be possible, the time increments of such quantities should be uniquely expressible in terms of the time increments of physical $w_i$. A more detailed description can be found in Yalim et al. (2017). 
For example, if $u>c_\mathrm{f}$, all physical quantities should be specified at the inner spherical boundary. If $c_\mathrm{A}<u<c_\mathrm{f}$, only 7 physical boundary conditions are possible, the remaining unknown variable should be found by solving the system of MHD equations.

Our model is designed to be driven by a variety of observational data primarily by Solar Dynamics Observatory~\citep{PTC12}/Helioseismic and Magnetic Imager~\citep{Schou12} (\textit{SDO}/HMI) synoptic/synchronic vector magnetogram data~\citep{Liu17}. The horizontal velocity components are obtained by applying the Differential Affine Velocity Estimator for Vector Magnetograms (DAVE4VM)~\citep{Schuck08,LZS13} and the time-distance helioseismology methods~\citep{Zhao12} to the HMI vector magnetogram data. In addition, our model can also be driven by line-of-sight (LOS) magnetogram data obtained by HMI, Solar and Heliospheric Observatory~\citep{DFP95}/Michelson Doppler Imager~\citep{Scherrer95} (\textit{SOHO}/MDI), National Solar Observatory/Global Oscillation Network Group (\textit{NSO}/GONG), and Wilcox Solar Observatory (\textit{WSO}). There is also a possibility of utilizing differential rotation~\citep{KHH93a} and meridional flow~\citep{KHH93b} formulae for horizontal velocity  at high latitudes where the time-distance helioseismology method data do not exist.

We solve the set of ideal MHD equations in the heliocentric, inertial or corotating frame of reference, using volumetric heating source terms to model solar wind acceleration by taking the 3D global magnetic field structure in the solar corona into account~\citep{Nakamizo09,Feng10}. They are written in corotating frame with the Sun, in terms of conservative variables, in conservation-law form as follows:

\begin{equation}
\label{eqGOV}
\frac{\partial}{\partial t}
\left(
\begin{array}{c}
\rho  \\
\rho \mathbf{v} \\
\mathbf{B} \\
E
\end{array}
\right) +
\nabla\cdot
\left(
\begin{array}{c}
\rho\mathbf{v} \\
\rho\mathbf{v}\mathbf{v} + \mathbf{I}(p+\frac{B^2}{8\pi})-\frac{\mathbf{B}\mathbf{B}}{4\pi} \\
\mathbf{v}\mathbf{B}-\mathbf{B}\mathbf{v} \\
(E+p+\frac{B^2}{8\pi})\mathbf{v}-\frac{\mathbf{B}}{4\pi}(\mathbf{v}\cdot\mathbf{B})
\end{array}
\right) =
\left(
\begin{array}{c}
0 \\
\rho[\mathbf{g}+(\mathbf{\Omega}\times\mathbf{r})\times\mathbf{\Omega}+2(\mathbf{v}\times\mathbf{\Omega})] + \mathbf{{S}}_{M} \\
0 \\
\rho\mathbf{v}\cdot[\mathbf{g}+(\mathbf{\Omega}\times\mathbf{r})\times\mathbf{\Omega}] + S_{E}
\end{array}
\right),
\end{equation} where $\rho$, $\mathbf{v}$, $\mathbf{B}$, $p$, $E$, and $\mathbf{g}$ are the density, velocity, magnetic field, thermal pressure, specific total energy of the plasma, and gravitational acceleration, respectively. The source terms in the momentum and energy conservation equations include the Coriolis and centrifugal forces, which are present only when the system is solved in a frame corotating with the Sun. Accordingly, $\mathbf{\Omega}$ and $\mathbf{r}$ correspond to the angular velocity of the Sun and position vector, respectively.

In order to model the solar wind acceleration, we introduce a volumetric heating source term, $S_{E}$, into the energy conservation equation, and the corresponding source term, $\mathbf{{S}}_{M}$, into the conservation of momentum equations~\citep{Nakamizo09,Feng10}. They are given as follows:

\begin{equation}
\label{eq2a}
S_{E}=\frac{Q_{0}}{f_{s}}\textrm{exp}\Big(-\frac{r}{L_{Q}}\Big)+\nabla\Big(\xi T^{2.5}\frac{\nabla T\cdot\mathbf{B}}{B^2}\Big)\cdot\mathbf{B}, \\
\end{equation} where the first term is an ad hoc heating function and the second term is a thermal conduction term of the Spitzer type, and

\begin{equation}
\label{eq2b}
\mathbf{S}_{M}=\frac{M_{0}}{f_{s}}\Big(\frac{r}{R_{\odot}}-1\Big)\textrm{exp}\Big(-\frac{r}{L_{M}}\Big),
\end{equation} where $T$ is the plasma temperature, $L_\mathrm{M}$, $L_\mathrm{Q}$, $M_0$, and $Q_0$ are the model constants given as $L_\mathrm{M}=L_Q=0.9 R_\odot$, $M_0=2.65\times 10^{-14}$ N $\mathrm{m}^{-3}$, and $Q_0=1.65\times 10^{-6}$ J $\mathrm{m}^{-3}$$\mathrm{s}^{-1}$. Additionally, $f_{s}$ is the expansion factor by which a magnetic flux tube expands in solid angle between its footpoint location on the photosphere and the source surface which is typically at $R_\mathrm{SS}=2.5R_\odot$~\citep{WS97}:

\begin{equation}
\label{eq3FS}
f_{s}=\frac{B(R_\odot)}{B(R_{SS})}\left(\frac{R_\odot}{R_{SS}}\right)^2.
\end{equation} 

These source terms take the coronal magnetic field topology into account by incorporating the expansion factor. 

The expansion factor is computed in every cell located between the inner boundary and the source surface according to the field-line tracing algorithm applied along the magnetic field lines presented in \citet{Cohen15}. The expansion factor depends on the evolution of the coronal magnetic field with distance from the Sun in the background solar wind solution. After a CME is introduced into the background solar wind, the expansion factor remains unchanged. Otherwise, the force imbalance created by the flux rope inserted into background solar wind results in unphysical results. Besides, the quasi-steady background solar wind solution that interacts with the CME has already a well-established coronal magnetic field structure and the expansion factor associated with it. We will later demonstrate a good overall agreement of the CME speed between our simulation results and observational data and in this way justify our treatment of the expansion factor.

We calculate the initial solution for magnetic field with a Potential Field Source Surface (PFSS) model~\citep{AN69,Schatten69} using either a spherical harmonics approach~\citep{Hoeksema84,WS92,SD03} or a finite difference method by incorporating the solution provided by the Finite Difference Iterative Potential-field Solver (FDIPS) code~\citep{TVH11}. For the rest of the plasma parameters, we compute the initial solution from Parker's isothermal solar wind model~\citep{Parker58}.

\subsection{Gibson-Low Flux rope model} \label{GL}

Solution to a GL flux rope is found by assuming balance of magnetic, pressure gradient and gravity forces. This can be written as  $(\nabla\times\mathbf{B})\times\mathbf{B}-\nabla p-\rho\mathbf{g}=0$, where $\mathbf{B}$ is the magnetic field, $p$ is the pressure, $\rho$ is the density, and $\mathbf{g}$ is the gravitational acceleration. To define a GL flux rope, we also use the Gauss's law of magnetism, i.e., $\nabla\cdot\mathbf{B}=0$. After deriving an analytical solution to $\mathbf{B}$ in the form of a spherical torus, a stretching transformation of $r\rightarrow r-a$ is used in spherical coordinates, where $r$ is the radial coordinate and $a$ is the stretching parameter. This results in a spherical torus of magnetic field lines being stretched into a tear drop shape. The analytical solution for a GL flux rope requires four parameters:
\begin{enumerate}
\item Flux rope radius ($r_0$): This is the radius of an initial GL spherical torus before stretching.
\item Flux rope height ($r_1$): This is the height of the center of the introduced spherical torus with respect to the center of the Sun before stretching.
\item Flux rope stretching parameter ($a$): This is the amount by which each part of the spherical torus is stretched towards the center of the Sun.
\item Flux rope field strength ($a_1$): This is a free parameter that controls the field strength in the flux rope being introduced. Plasma pressure inside the rope is proportional to $a_1^2$ due to the condition of pressure balance assumed in this solution.   
\end{enumerate}

\begin{figure}[p!]
\vspace{-1.5cm}
\center
\begin{tabular}{c c}  
\vspace{-0.7cm}
\includegraphics[scale=0.1,angle=0,width=6cm,keepaspectratio]{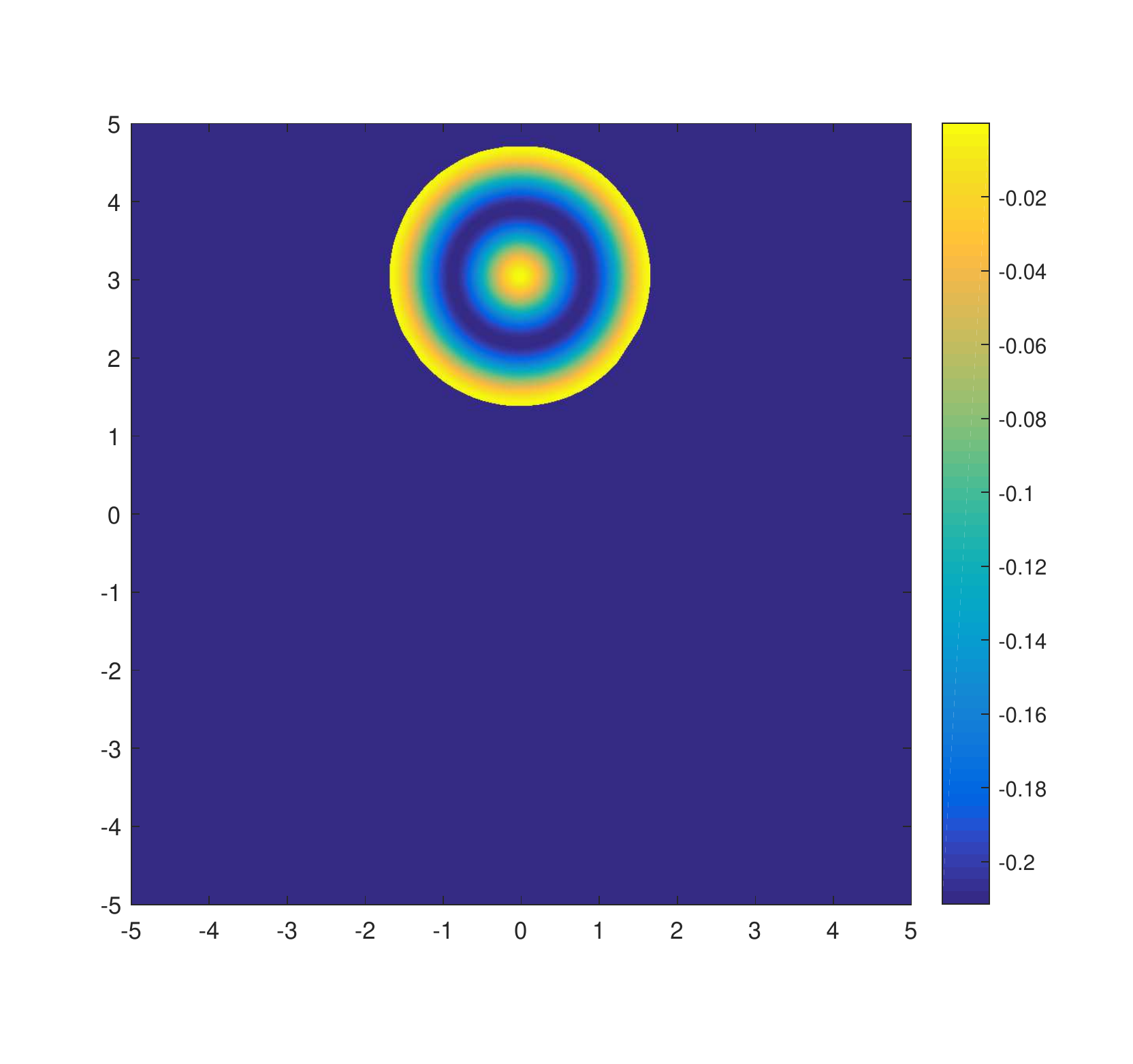}
\hspace{-0.5cm}
\includegraphics[scale=0.1,angle=0,width=6cm,keepaspectratio]{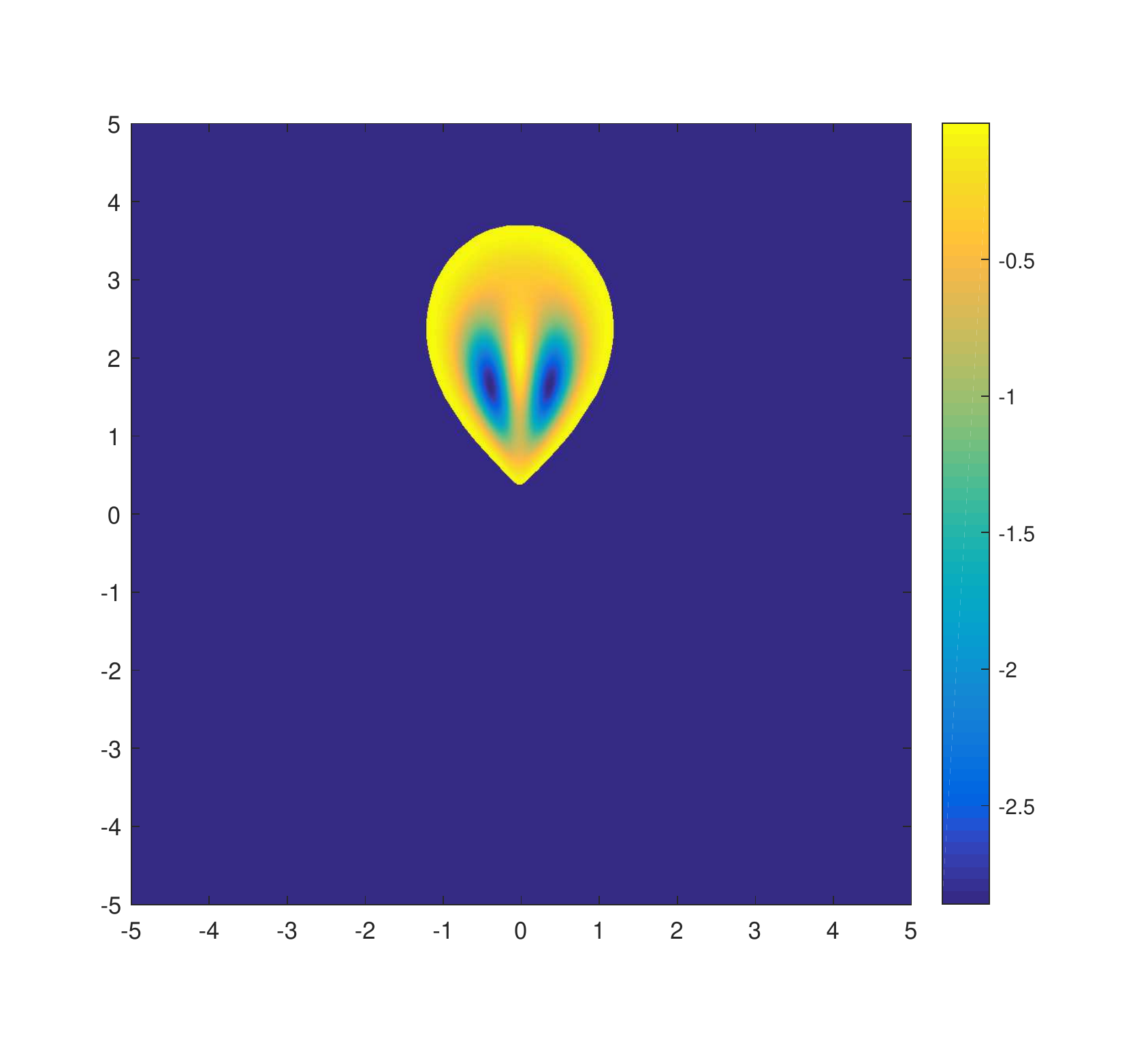} \\
\vspace{-0.7cm}
\includegraphics[scale=0.1,angle=0,width=6cm,keepaspectratio]{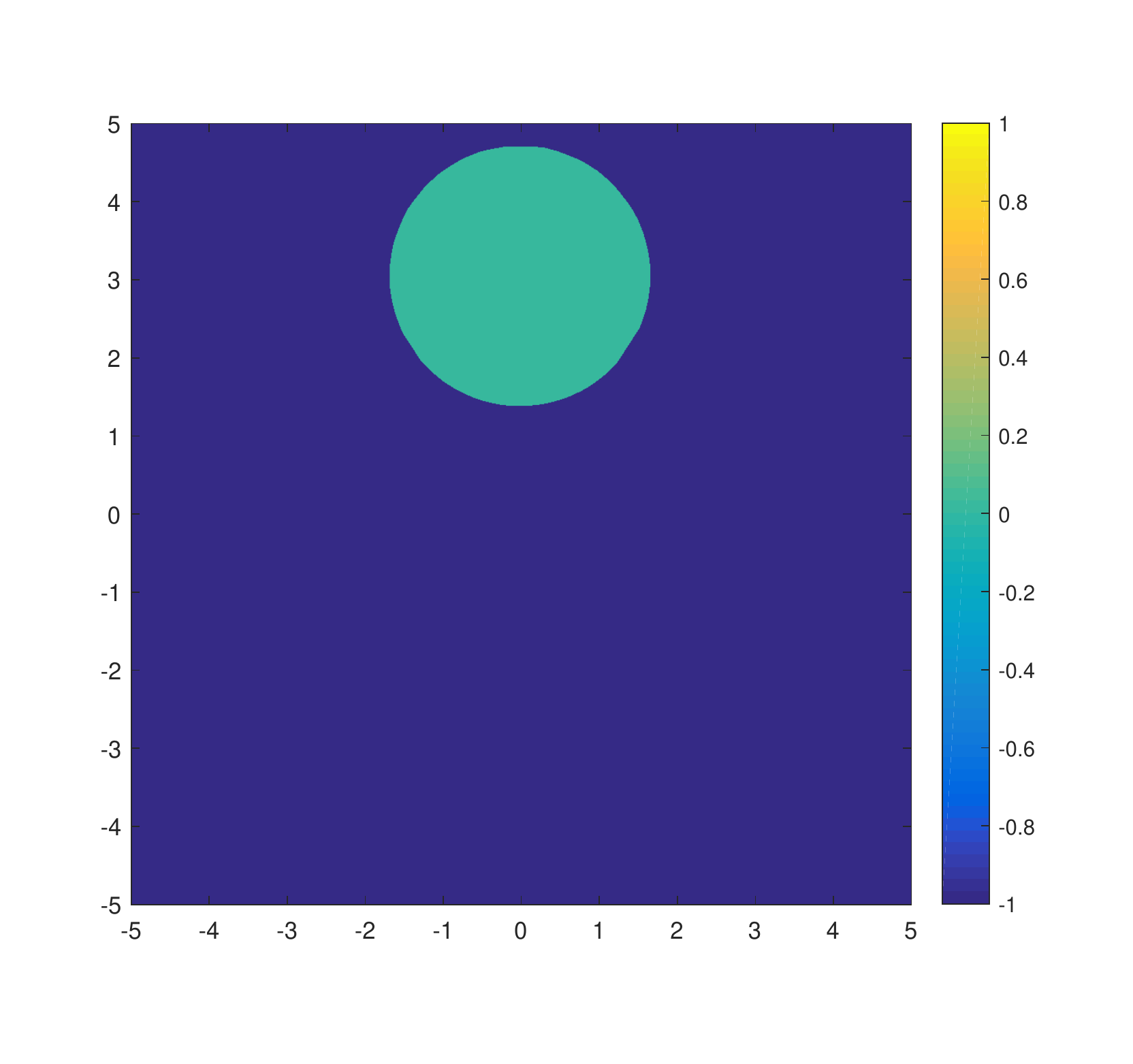}
\hspace{-0.5cm}
\includegraphics[scale=0.1,angle=0,width=6cm,keepaspectratio]{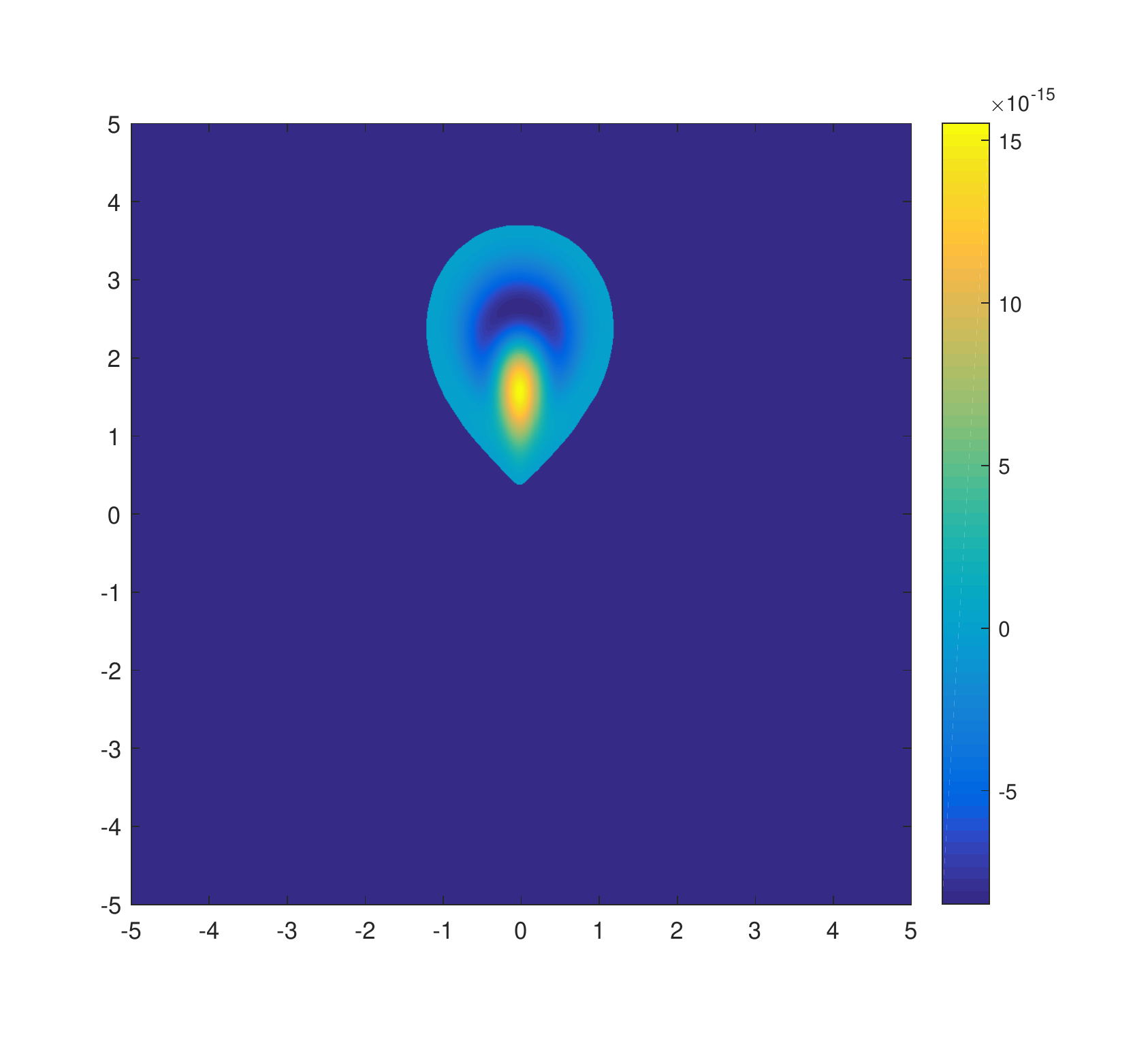}\\
\vspace{-0.7cm}
\includegraphics[scale=0.1,angle=0,width=6cm,keepaspectratio]{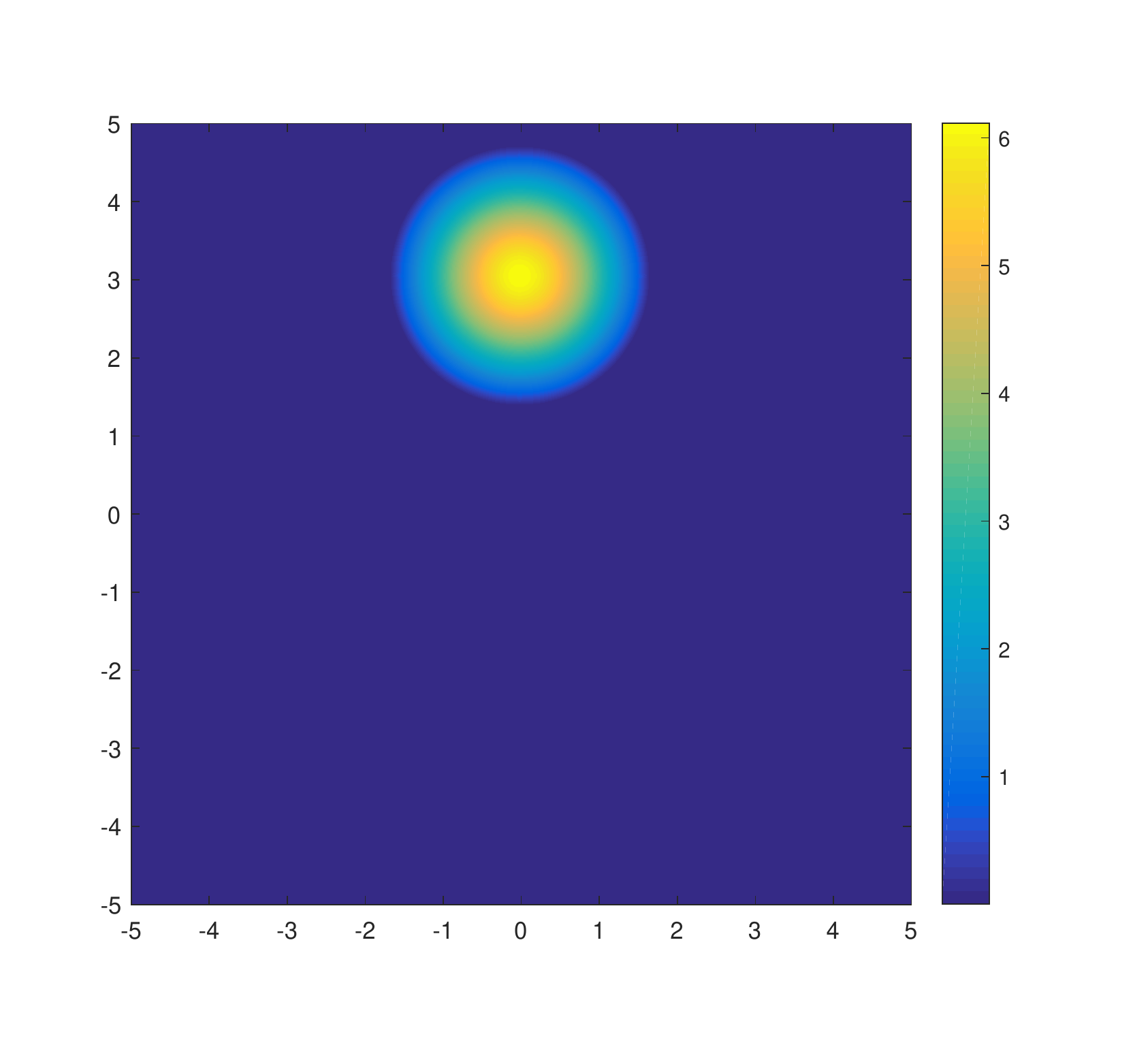}
\hspace{-0.5cm}
\includegraphics[scale=0.1,angle=0,width=6cm,keepaspectratio]{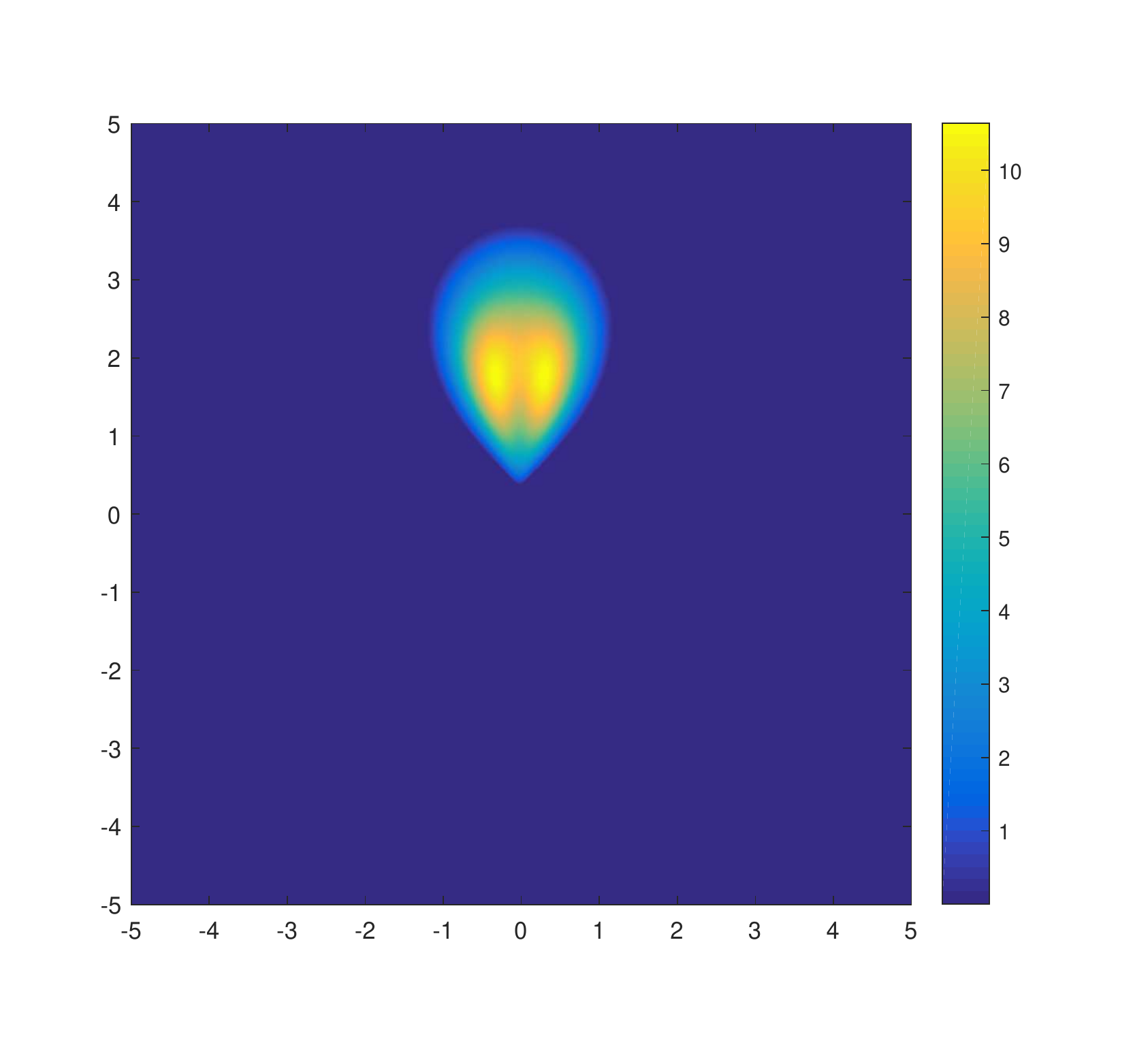}\\
\hspace{-0.7cm}
\includegraphics[scale=0.1,angle=0,width=5.6cm,keepaspectratio]{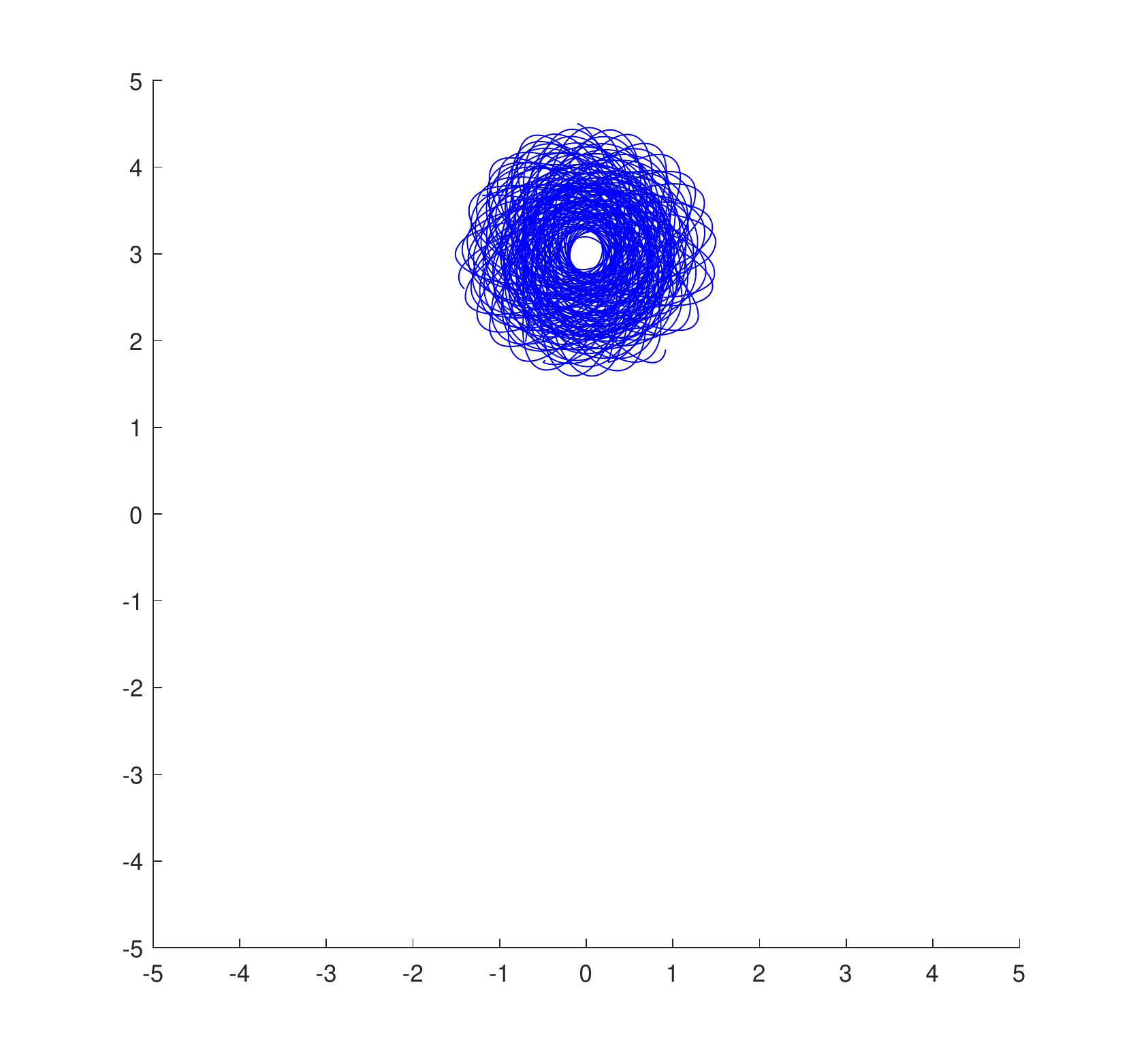} 
\hspace{-0.2cm}
\includegraphics[scale=0.1,angle=0,width=5.6cm,keepaspectratio]{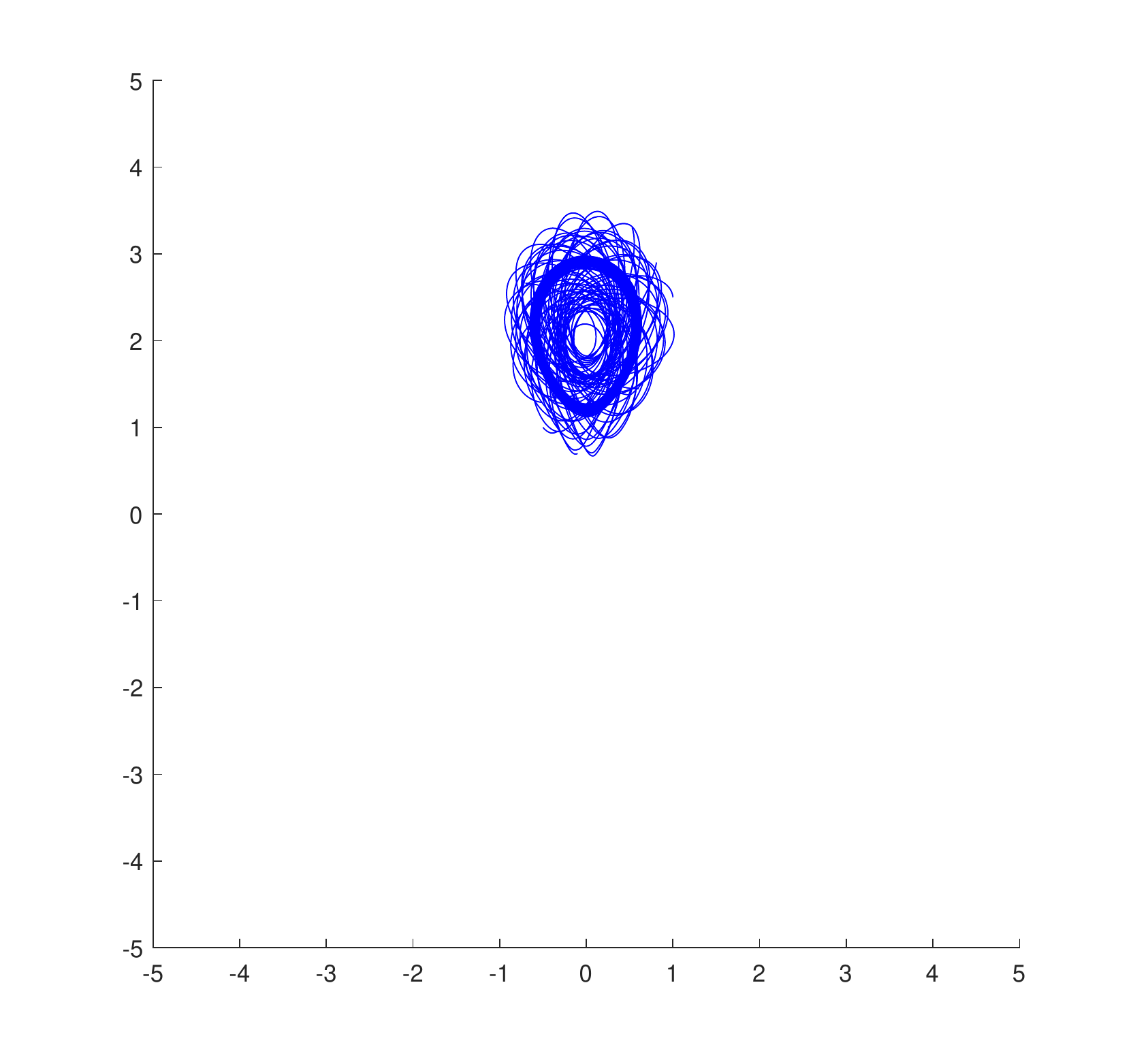} 

\end{tabular}

\caption{(From \textit{top} to \textit{bottom}) Pressure ($dyne/cm^2$), density ($g/cm^3$), magnetic field magnitude ($G$) and magnetic filed lines in unstretched(\textit{left}) and stretched(\textit{right}) GL torus. All horizontal and vertical axis are in $R_{\odot}$. We used $r_0=1.67$, $r_1=3.03$, $a=1.01$ and $a_1=0.23$ in these figures. (0,0) coordinate represents solar center.}
\label{GL0}
\end{figure}

Figure \ref{GL0} shows pressure, density, magnetic field magnitude, and magnetic field lines in the plane containing the centroidal axis of a spherical torus before and after the stretching operation. Notice that density is introduced only after stretching, since no force is associated with it before stretching.

\subsection{Data-Constrained CME Model Using Graduated Cylindrical Shell Method} \label{GCS}

\begin{figure}[ht]
\center
\includegraphics[scale=0.4,angle=0,width=15cm,height=15cm,keepaspectratio]{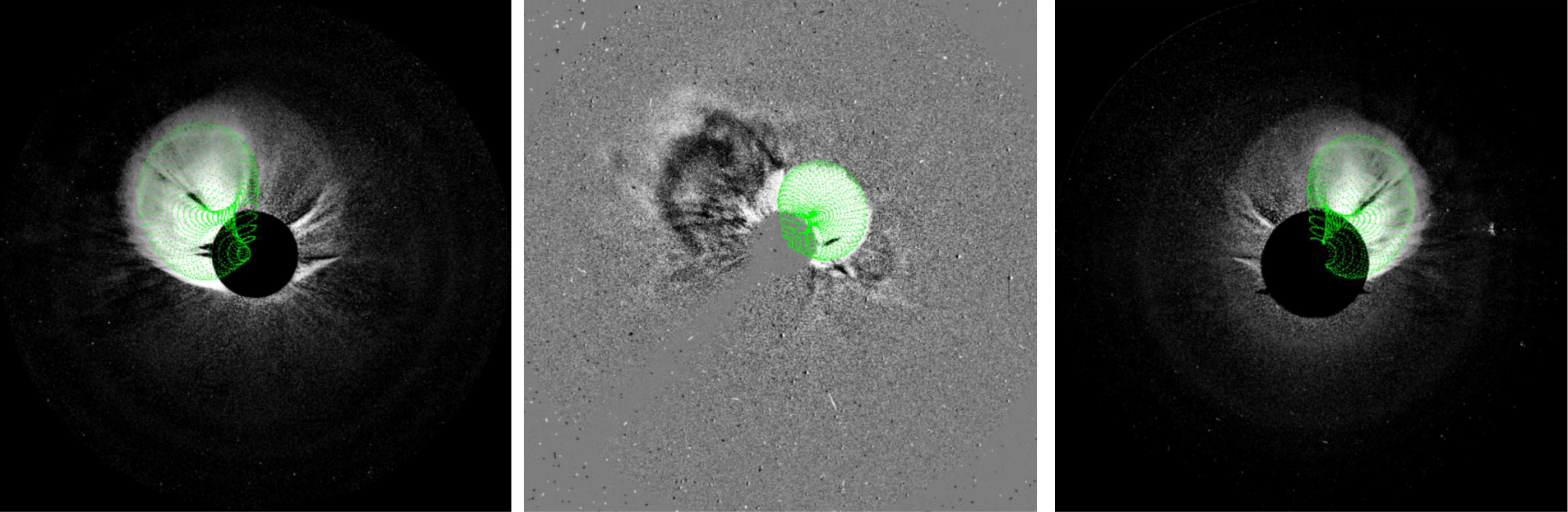} 
\caption{GCS fitting of a CME using three viewpoints: (left and right panels) SECCHI/Cor2 onboard $\textit{STEREO A \& B}$ respectively, and (middle panel) LASCO/C3 onboard $\textit{SOHO}$.}
\label{fig1}
\end{figure}

\begin{figure}[ht]
\center
\includegraphics[width=.5\textwidth, angle=0]{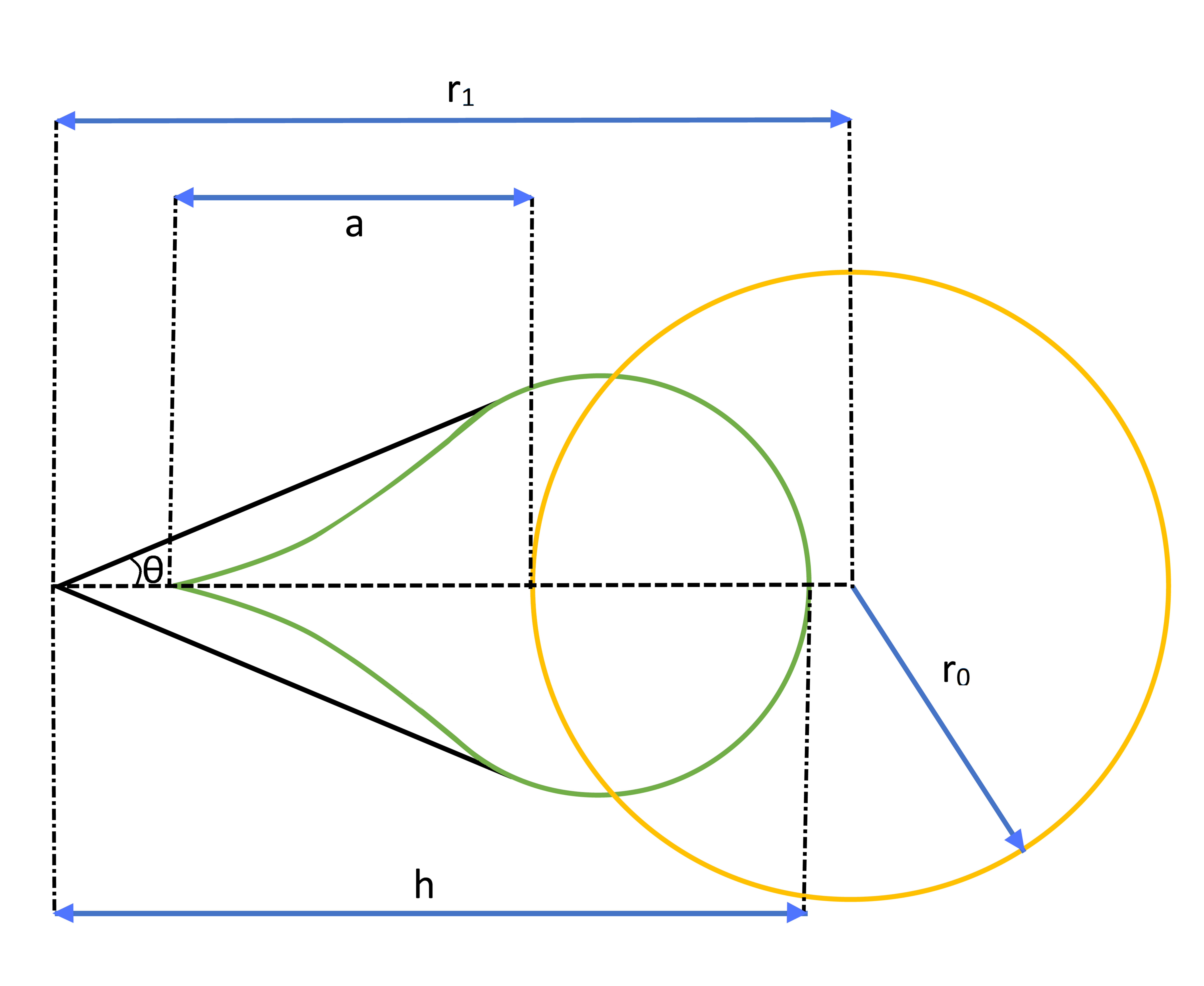} 
\caption{A diagram showing the GL sphere (yellow), its stretched tear drop shape (green), and GCS fit outline (black).}
\label{fig2}
\end{figure}

We utilize the Sun-Earth Connection Coronal and Heliospheric Investigation (SECCHI)/Cor1/Cor2 ~\citep{Howard08} coronagraph image data from \textit{STEREO A \& B}~\citep{Kaiser08} and Large Angle Spectroscopic Coronagraph (LASCO)/C2/C3~\citep{Brueckner95} data from \textit{SOHO} as observational data to constrain the GL flux rope parameters. We apply the GCS method to find the height ($h$), direction and half angle ($\theta$) (from the central axis to the outer edge) of the CME as shown in Figs.~\ref{fig1} and~\ref{fig2}. GCS fitting is a visual fitting tool where three viewpoints of a CME from \textit{STEREO A \& B} and \textit{SOHO} coronagraphs are used to fit the flux rope structure with conical legs and curved fronts over a CME. The GCS method was implemented in IDL using the \textit{rtsccguicloud} program~\citep{THV06}. The size parameters of a GL flux rope $r_0$, $r_1$, and $a$ can be approximately related to the GCS size parameters according to the geometry shown in Fig.~\ref{fig2}.

We work under the assumption that $a=r_{1}/3$ and the front edge of tear drop shape roughly matches the front end of GCS shape. In fact, by comparing the curved fronts of the tear drop and GCS shapes, we find that if we vary $r_0$ from $0.4$  to 2 $R_{\odot}$ and $r_1$ from 1.5 to 5 $R_{\odot}$, the maximum distance between the two shapes is always less than 5\% of $r_1$. Therefore, the two shapes coincide very well. Therefore,
\begin{eqnarray}
h+a=r_1+r_0,\\
h-r_0=\frac{2}{3}r_1,\\
h-r_1 \,\sin\theta=\frac{2}{3}r_1.
\label{eq1}
\end{eqnarray}

This gives us:

\begin{equation}
r_1=\frac{h}{2/3\,+\sin\theta};\quad a=\frac{h}{2+3\sin\theta}; \quad r_0=\frac{h\,\sin\theta}{2/3\,+\sin\theta}.
\label{eq2}
\end{equation}

We notice that this approach constrains us to using the relation $r_0=r_1\sin\theta$. However, $r_1$ and $r_0$ are independent parameters in GL analytical formulae. Therefore, the dependence of $r_0$ on $r_1$ is only due to the observational limitations.

The remaining GL parameter (i.e. magnetic field strength, $a_1$) can not be determined from observations directly. Therefore, we perform a parametric study to find an expression for $a_1$ in terms of $r_0,\, r_1$, the average simulated solar wind pressure above the erupting region, $P_\mathrm{avg}$, and speed of a CME, $V_\mathrm{CME}$. The latter can be found by applying linear fitting to the height vs. time data from the GCS method. In order to calculate $P_\mathrm{avg}$, we find average pressure in simulated solar wind in $\pm 30^\circ$ latitude and longitude from inner boundary to 10 $R_\odot$.  

\subsubsection{Parametric Study}

We follow the method used by \citet{Jin17a} to perform the parametric study. Here, we check the effect of changes in the input GL flux rope parameters on the CME speed. In contrast to \citet{Jin17a}, we additionally allow variations in $r_1$. There is also a possibility of using GCS size parameters for parametric study but, as we will show below, using GL size parameters gives results in form of simple linear functions. To perform the parametric study, we need to select a magnetogram with multiple active regions. At least one of the active regions should have ejected a CME in such a direction that the CME parameters can be easily determined by the GCS method. In this study, we select the HMI LOS magnetogram from 7 March 2011 06:00 UT, in which one of the active regions numbered AR11164 produced a fast CME that occurred on 7 March 2011 at 20:00 UT. We determine the size parameters of the GL flux rope corresponding to this CME as $r_0=1.68$, $r_1=3.03$ and $a=1.01$ using the GCS method. 

\begin{figure}[ht]
\center
\includegraphics[width=.7\textwidth, angle=0]{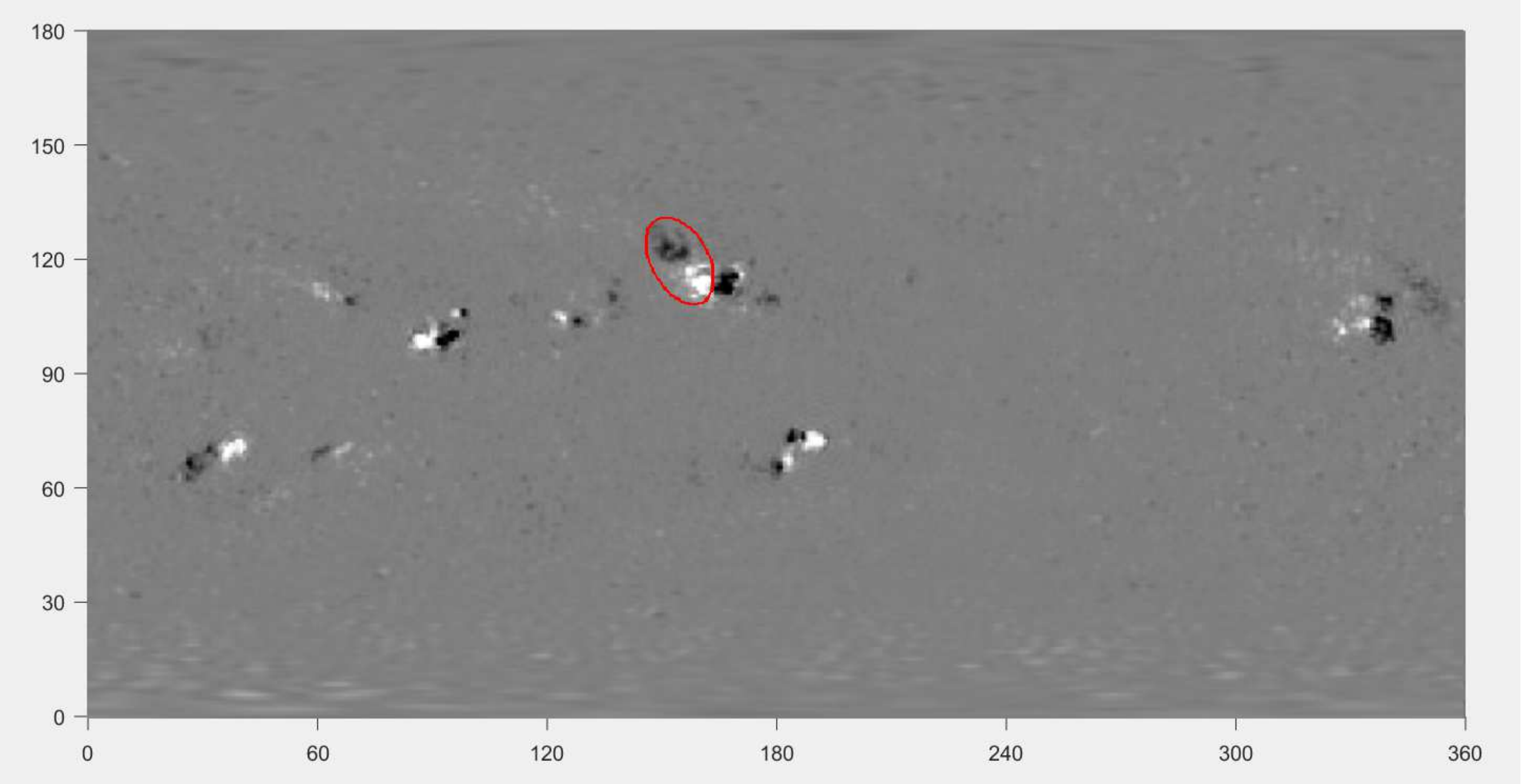} 
\caption{HMI LOS magnetogram obtained on 7 March 2011 at 06:00 UT with the source active region from which the CME erupted is indicated in red.}
\label{figMag}
\end{figure}

\begin{figure}[ht]
\center
\includegraphics[width=.9\textwidth, angle=0]{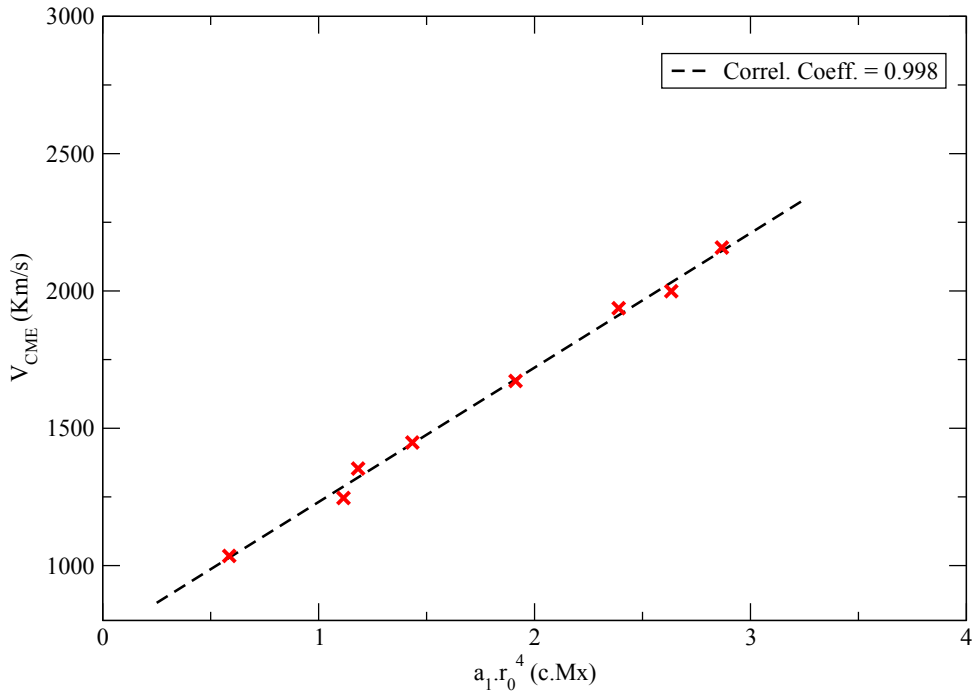} 
\caption{Variation of CME speed with poloidal flux ($\phi\propto a_1 r_0^4$).}
\label{fig4}
\end{figure}

\begin{figure}[ht]
\center
\includegraphics[width=0.9\textwidth, angle=0]{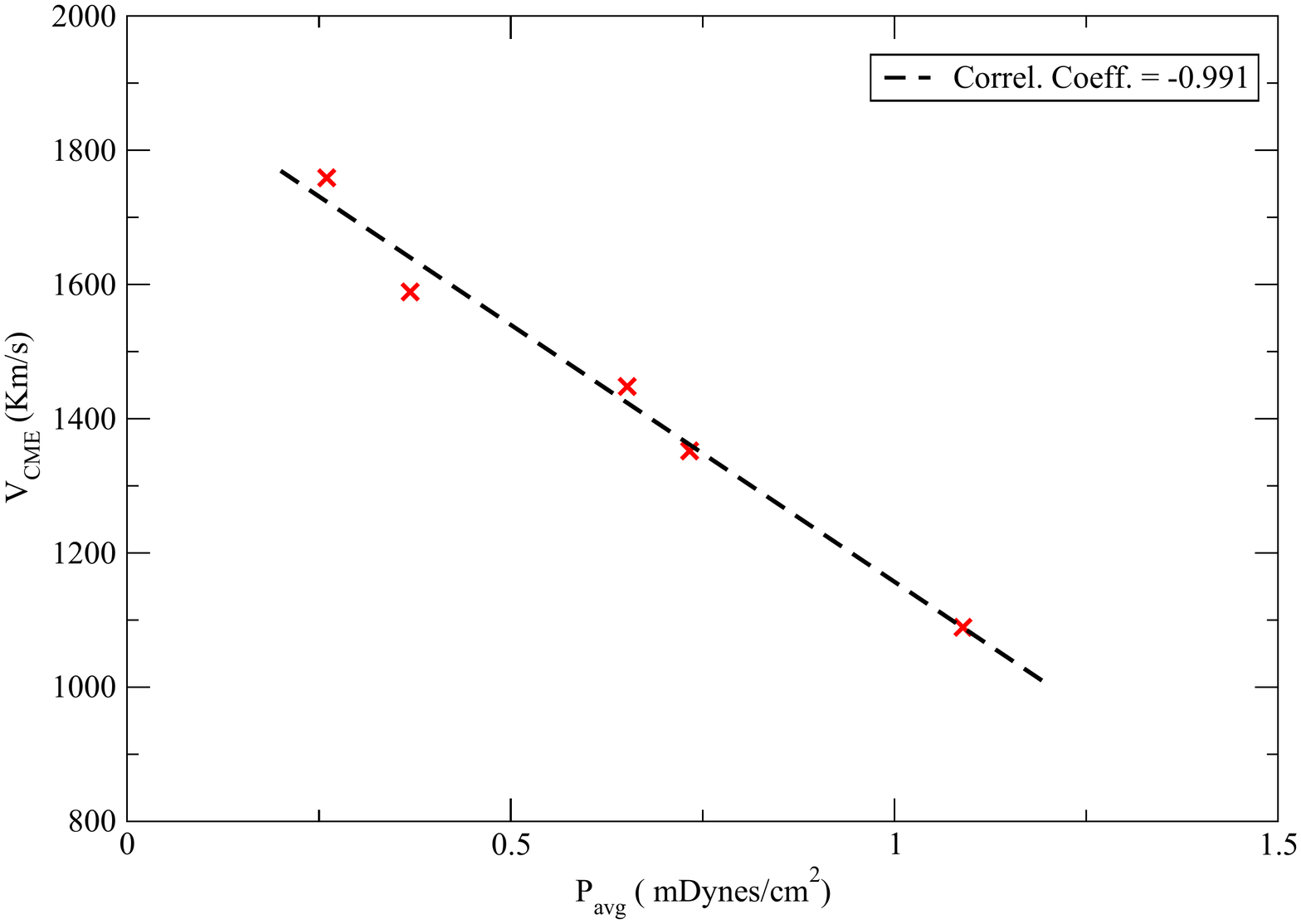} 
\caption{Variation of CME speed with $P_\mathrm{avg}$.}
\label{fig5}
\end{figure}

\begin{figure}[ht]
\center
\includegraphics[width=.9\textwidth, angle=0]{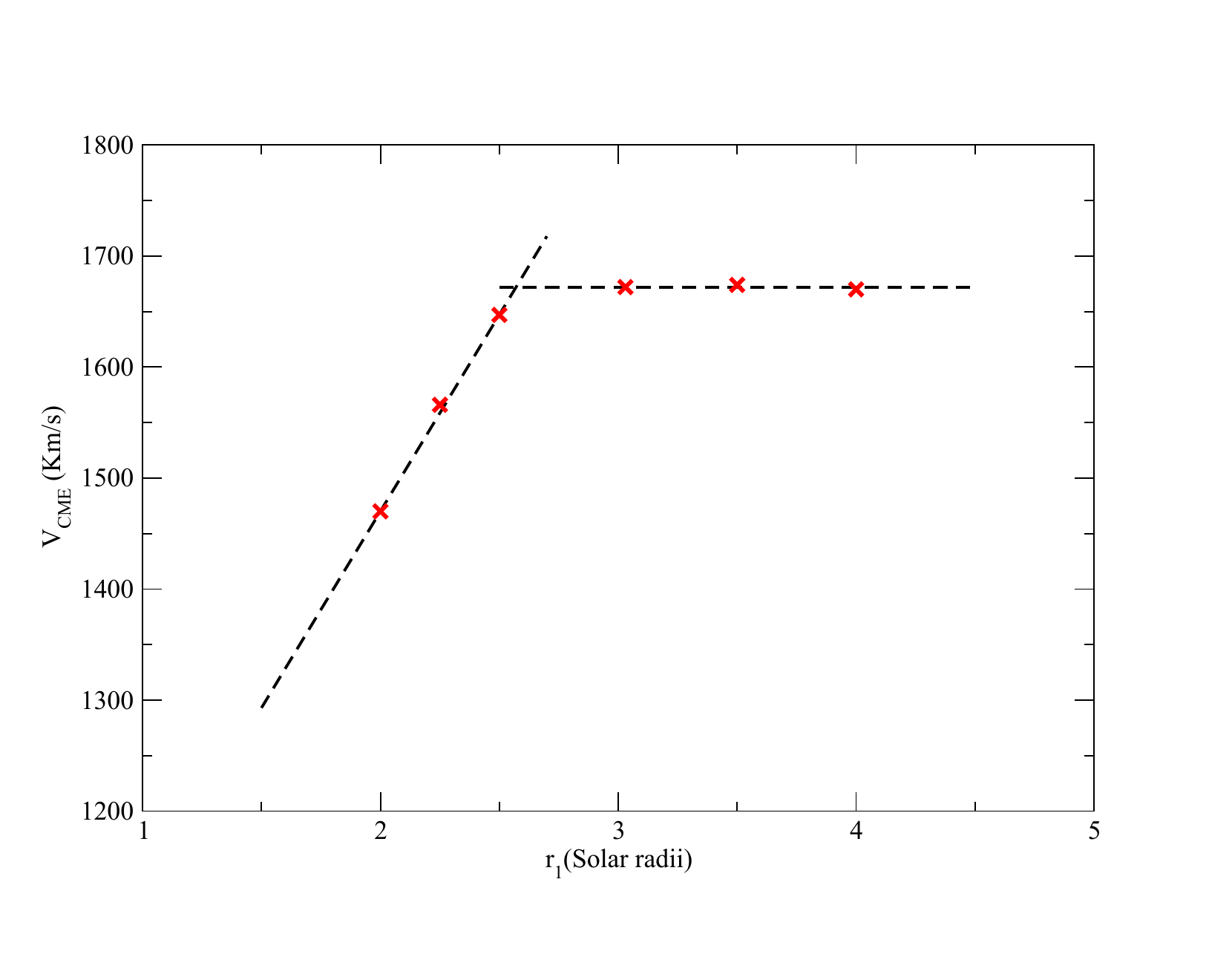} 
\caption{Variation of CME speed with $r_1$.}
\label{fig3}
\end{figure}

We perform our parametric study in three steps. First, a GL flux rope with $r_0=1.68$, $a_1=0.24$, and varying $r_1$ is kept on the source active region (AR11164) and the simulated CME speed is calculated (see Fig.~\ref{fig3}). Then, we fix $r_1=3.03$ and the poloidal flux, $\phi$, is varied by changing $r_0$ and $a_1$ while still keeping the flux rope at the same source active region (see Fig.~\ref{fig4}). Poloidal flux of a GL flux rope can be determined by integrating the  magnitude of poloidal magnetic field component over the surface perpendicular to the polar axis of GL spherical torus. It can be shown that $\phi\propto a_1 r_0^4$ \citep{Jin17a}.  Finally, we place the same flux rope with parameters $r_0=1.68$, $a_1=0.18$ and $r_1=3.03$ over different active regions with different $P_\mathrm{avg}$'s and determine the variation in the simulated CME speed (see Fig.~\ref{fig5}). 

We combine all these steps to derive an expression for $a_1$ as follows:

\begin{equation}
V_\mathrm{CME}=f_1(\phi) \cdot f_2(P_{avg}) \cdot f_3(r_1).
\label{eq3}
\end{equation}

The parametric study shows that $f_1$ and $f_2$ are linear functions whereas $f_3$ is linear for $r_1<2.6$ and constant for $r_1>=2.6$. There is an explanation for the latter behavior. When we keep the stretched GL flux rope closer to the Sun, most of its lower part resides under solar surface and full energy of the GL flux rope is therefore not injected into the background solar wind. We also note that \citet{Jin17a} use active region magnetic field strength $B_r$ instead of $P_\mathrm{avg}$ to differentiate between different locations where flux rope is kept initially. We find that $P_\mathrm{avg}$ shows much better correlation with $V_\mathrm{CME}$ than $B_r$. 
 
Keeping the above in mind, we can write out:

\begin{equation}
V_\mathrm{CME}=\begin{cases}
(c_1a_1r_0^4+c_2)\cdot(c_3P_{avg}+c_4)\cdot(c_5r_1+c_6) &r_1<2.6 \\
(c_1a_1r_0^4+c_2)\cdot(c_3P_{avg}+c_4) &r_1\geq 2.6
\end{cases}
\label{eq4}
\end{equation}

Now, we use non-linear multi-variable regression on all the CME runs in the parametric study to find the fitting constants. The results are given in Table~\ref{tab1}. Finally, the expression for $a_1$ can be written as follows:

\begin{equation}
a_1=\begin{cases}
\frac{1}{c_1r_0^4}\cdot\Big(\frac{V_\mathrm{CME}}{(c_3P_{avg}+c_4)\cdot(c_5r_1+c_6)}-c_2\Big) &r_1<2.6 \\
\frac{1}{c_1r_0^4}\cdot\Big(\frac{V_\mathrm{CME}}{(c_3P_{avg}+c_4)}-c_2\Big) &r_1\geq 2.6
\end{cases}
\label{eq5}
\end{equation}

\begin{table}[]
\centering
\begin{tabular}{|l|l|l|l|l|l|l|}
\hline
$r_1$    & $c_1$    & $c_2$    & $c_3$    & $c_4$    & $c_5$    & $c_6$    \\ \hline
$<$ 2.6 & 3.849 & 5.831 & -6.018 & 15.112 & 2.783 & 6.009 \\ \hline
$\geq$ 2.6 & 13.018 & 19.721 & -20.354 & 51.112 &   &   \\ \hline
\end{tabular}
\caption{Parameters used in the expression of $a_1$.}
\label{tab1}
\end{table}

\section{Simulation Results} \label{res}

\begin{figure}[ht]
\center
\includegraphics[width=1\textwidth, angle=0]{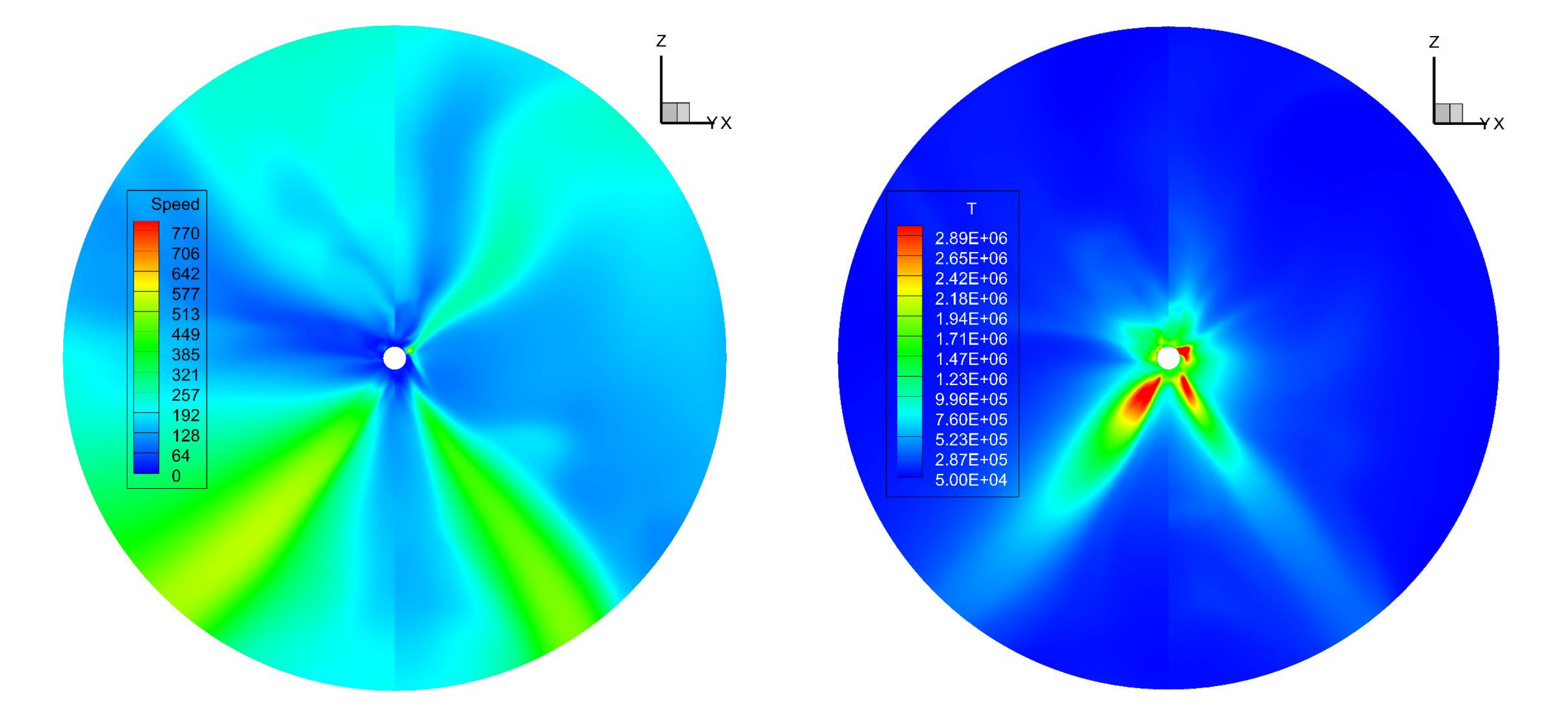} 
\caption{Solar wind background simulated using HMI LOS magnetogram of 7 March 2011 06:00 UT: (left) speed contours (km/s); (right) temperature contours (K). Background is shown in the plane in which flux rope is introduced.}
\label{fig5a}
\end{figure}

In this section, we show the results related to our simulation of the eruption of the fast CME that occurred on 7 March 2011 at 20:00 UT. The background solar wind solution is obtained by relaxing the initial PFSS magnetic field distribution to steady state using our data-driven MHD global solar corona model. We computed the initial conditions for magnetic field corresponding to the simulation made to obtain the background solar wind solution from the PFSS model by using the spherical harmonics coefficients corresponding to the HMI LOS magnetogram on 7 March 2011 obtained from the \textit{pfss\_viewer} program on IDL SolarSoft. The initial conditions for the remaining hydrodynamic plasma variables were obtained from Parker's isothermal solar wind model. We used the TVD, finite volume Rusanov scheme~\citep{KPS01} to compute the numerical fluxes and the forward Euler scheme for time integration. In order to satisfy the solenoidal constraint, we applied Powell's source term method~\citep{Powell99}. Our computational domain size is 1.03$R_{\odot}\leq$ r $\leq 30R_{\odot}$, $0\leq\phi\leq 2\pi$, $0\leq\theta\leq\pi$ and grid size is 180$\times$240$\times$120 in r, $\phi$ and $\theta$ directions, respectively. We perform all simulations in the frame corotating with the Sun. MSFLUKSS provides us with parallel implementation of the numerical methods. At the inner boundary of the computational domain which is located at the lower corona, we applied the radial magnetic field derived from the HMI LOS magnetogram data and the differential rotation~\citep{KHH93a} and meridional flow~\citep{KHH93b} formulae for the horizontal velocity components at the ghost cell centers. We kept density and temperature constant as $n=1.5 \times 10^{8} \,\textrm{cm}^{-3}$ and $T=1.3 \times 10^{6}$ K, respectively. The radial velocity component is imposed to be zero at the boundary surface. The transverse magnetic field components are extrapolated from the domain into the ghost cells. At the outer boundary of the domain which is located beyond the critical point, the plasma flow is superfast magnetosonic, so no boundary conditions are required. The computational domain, grid size, numerical methods and boundary conditions are the same for all the runs performed. Figure~\ref{fig5a} shows the background through which we propagate the CME.

\begin{figure}[ht]
\center
\includegraphics[width=.8\textwidth, angle=0]{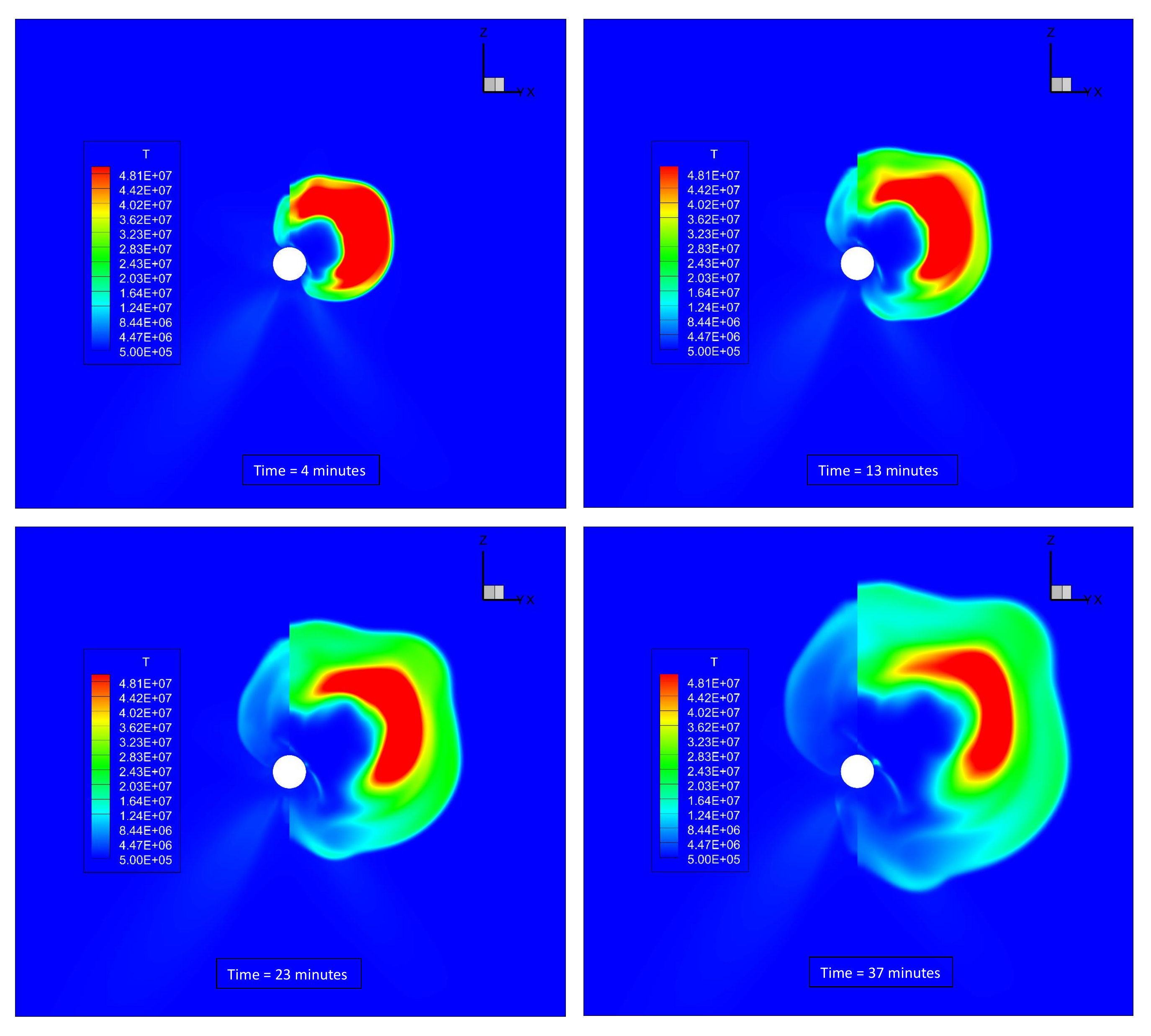} 
\caption{Time evolution of the CME shown using temperature contours.}
\label{fig5b}
\end{figure}

\begin{figure}[ht]
\center
\includegraphics[width=.8\textwidth, angle=0]{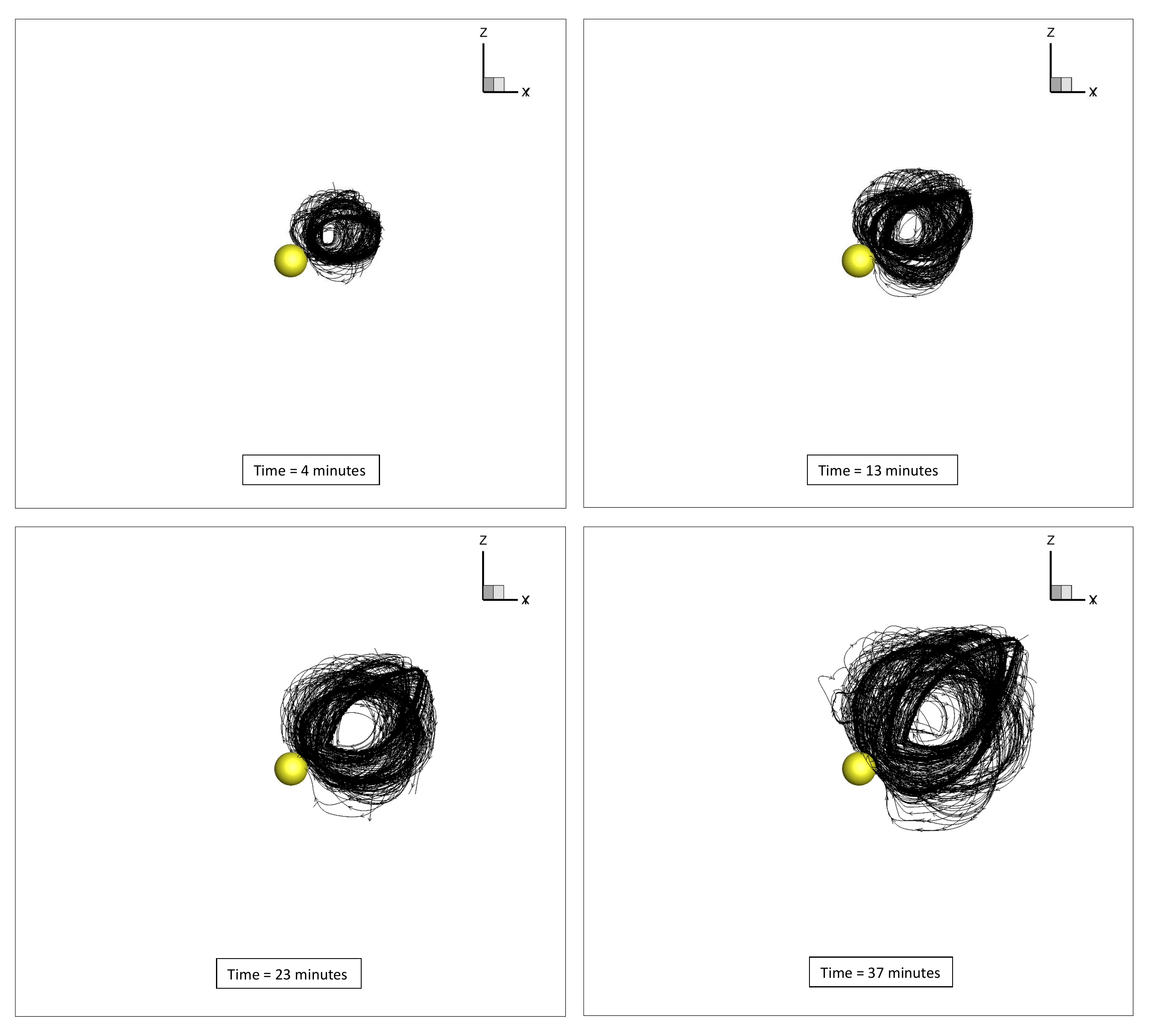} 
\caption{Time evolution of the CME shown using magnetic field lines of flux rope.}
\label{fig5c}
\end{figure}

\begin{figure}[ht]
\center
\includegraphics[width=.8\textwidth, angle=0]{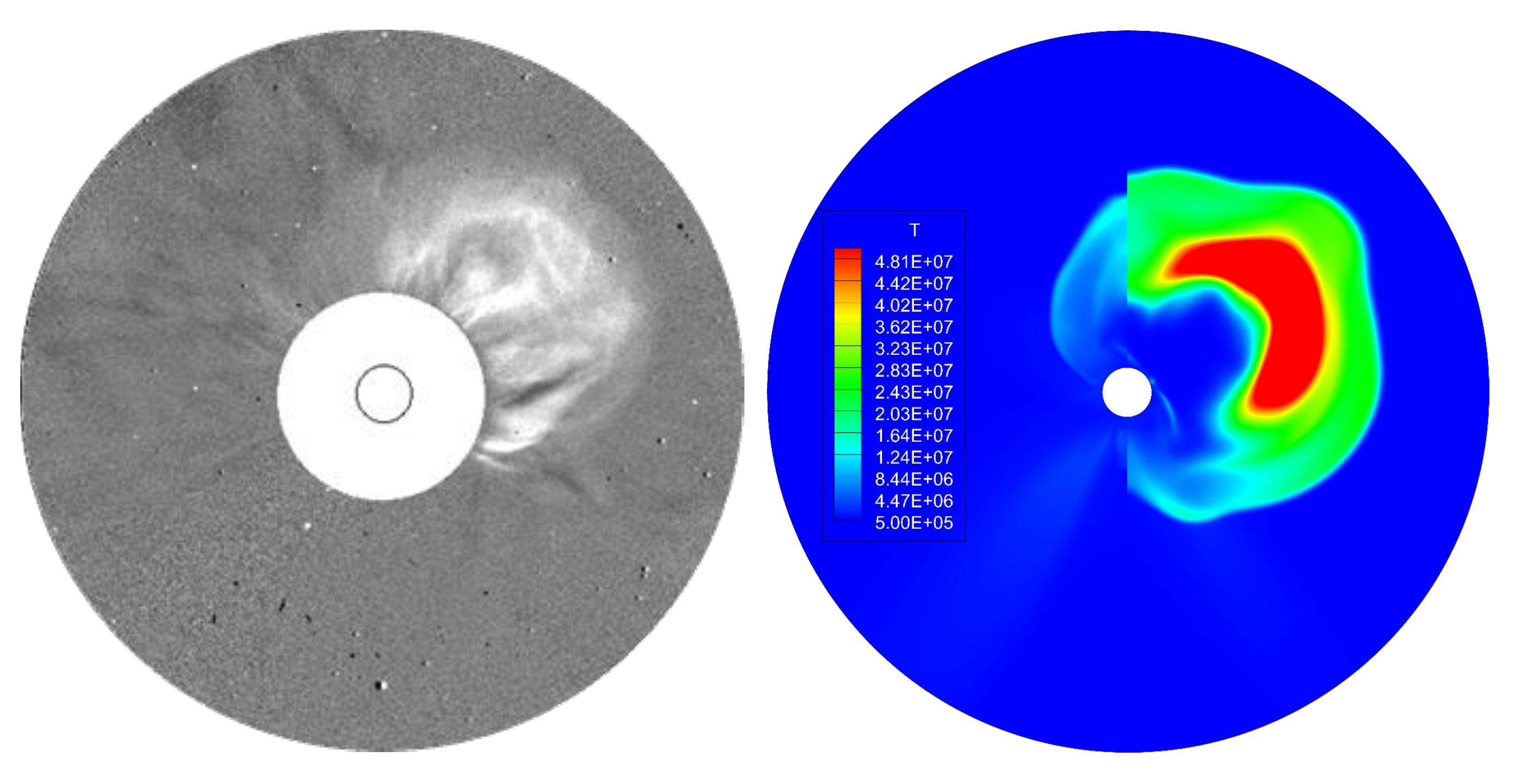} 
\caption{Comparison of CME shapes: (left panel) LASCO/C3 coronagraph difference image; (right panel) temperature contours obtained from the simulation.}
\label{fig6}
\end{figure}

\begin{figure}[ht]
\center
\includegraphics[width=.8\textwidth, angle=0]{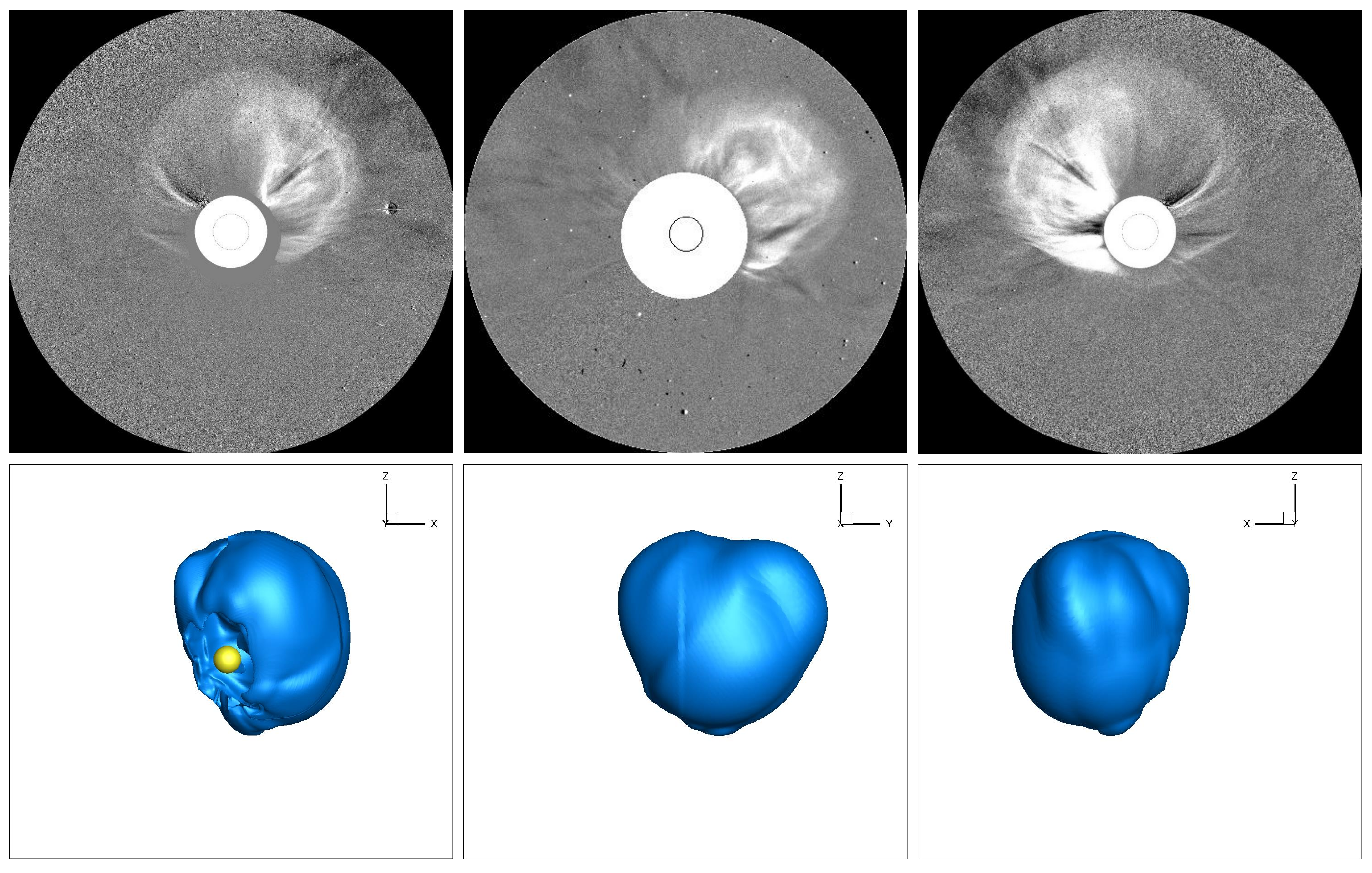}
\caption{Comparison of CME shapes from points of view of \textit{SOHO}, \textit{STEREO A \& B}: (upper row) From left to right: Cor2 (\textit{STEREO B}), LASCO/C3, Cor2 (\textit{STEREO A}); (lower row) Temperature Iso-surface indicating the CME shape in the same orientation as the corresponding image in the upper row and having same scales.}
\label{fig7}
\end{figure}

We use $V_\mathrm{CME}$=2125 km/s for the fast CME that occurred on 7 March 2011 as given by the SOHO CME catalog~\citep{Gopalswamy09}. We also found $r_0=1.68$, $r_1=3.03$ and $a=1.01$ from the GCS method for this CME. The pressure  $P_\mathrm{avg}$ is found to be 0.652 mdyne/cm$^2$. Using these values and the calculated coefficients in Table~\ref{tab1} in Eq.~\ref{eq5}, we find $a_1= 0.35$. Running our simulation with this value of $a_1$ and the GL size parameters, we find the simulated speed to be 2140 km/s which is very close to the actual speed. Figure \ref{fig5b} shows the time evolution of simulated CME using temperature contours whereas Fig. \ref{fig5c} shows the same time evolution using magnetic flux rope structure of the CME. Figures \ref{fig6} and \ref{fig7} shows the comparison of the simulated CME shape with the coronagraph observations. The shape of CME is approximated by an iso-surface of the temperature.

\begin{figure}[ht]
\center
\includegraphics[width=.8\textwidth, angle=0]{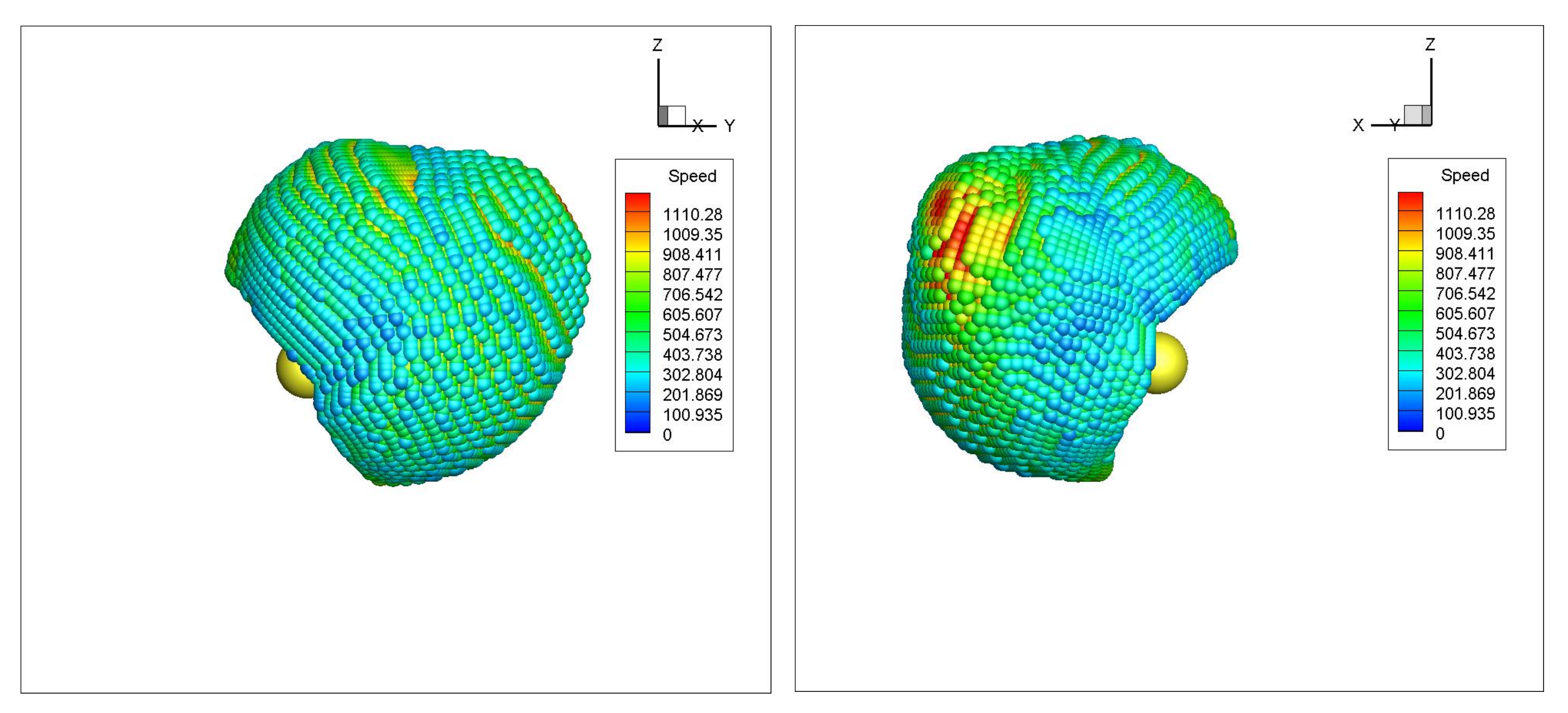}
\caption{Shock surface in front of the simulated CME from two different viewpoints at 23 minutes after eruption.}
\label{fig7a}
\end{figure}

Figure \ref{fig7a} shows the shock surface propagating in front of the CME. Its surface is colored according to speed values. The shock surface is found by locating the jump in entropy along radial direction with a resolution of 2$^{\circ}$ in latitude and longitude. Shock properties derived from our simulation can be used to model SEP events. \citet{Hu18} have recently used CME driven shocks to model SEP acceleration using their improved Particle Acceleration and Transport in the Heliosphere (iPATH) model. However, they note that more realistic treatment of CMEs in simulations, like using flux rope models, can enhance the accuracy of their results and better understand the SEP events. 

\begin{figure}[ht]
\center
\includegraphics[width=.8\textwidth, angle=0]{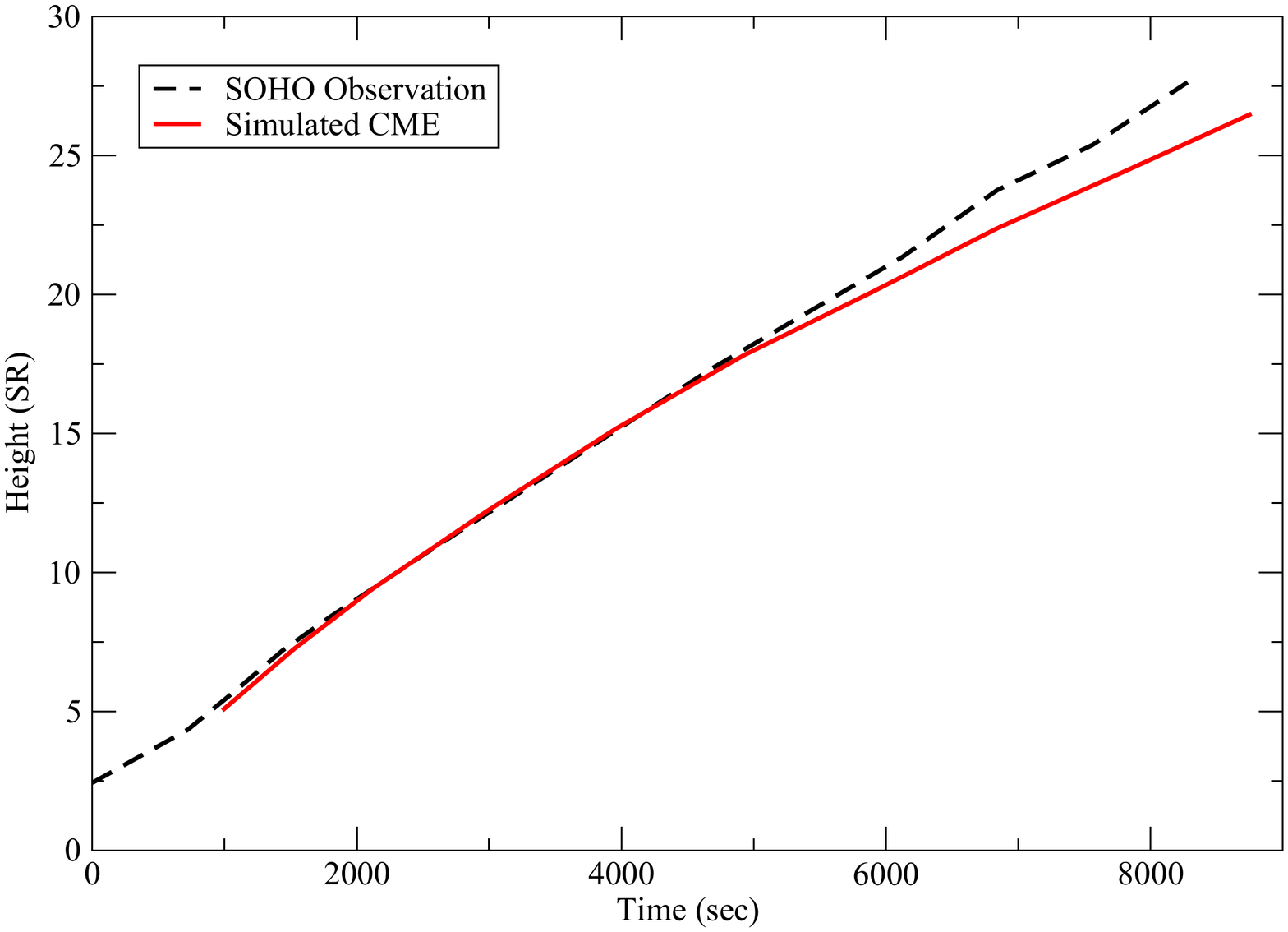} 
\caption{Comparison of height vs.time graphs between LASCO/C3 observations and simulation results.}
\label{fig8}
\end{figure}

Finally, Fig.~\ref{fig8} shows the agreement between the height vs. time graphs obtained from the LASCO/C3 observations and the simulation results. 



\section{Conclusions} \label{conc}

In this paper, we presented a data-constrained CME model which is based on the GL flux rope approach and uses the GCS method. Our CME model is complementary to the model described in~\citet{Jin17b} and has certain advantages over it. We determine the GL flux rope size parameters more accurately because of the application of the GCS method to \textit{SOHO}/LASCO/C2/C3 and \textit{STEREO A \& B}/SECCHI/Cor1/Cor2 coronagraph image data. Thus, we do not impose excessive energy in the initial flux rope configuration thereby avoiding excessive heating and acceleration of the flux rope. Determining the size parameters from the GCS method results in a realistic initial flux rope size in agreement with the observations, which leads to correct CME speed and acceleration. 

These results do not imply that our CME model is better than models involving energy buildup before eruption~\citep[e.g.][]{Titov18,ACA14}. However, due to its simplicity, our approach is less time consuming. Besides, it has obvious advantages over the ``blob" and ``cone" models because of a more realistic treatment of magnetic field.

Now when it is demonstrated that our data-constrained CME generation model works in the solar corona, we will propagate the same CME through the inner heliosphere and compare our simulation results with the near-Earth spacecraft data at 1 AU. We also plan to investigate CME-CME interactions in the future following a simulation approach.   


The authors acknowledge the support from the NASA project NNX14AF41G and NSF SHINE grant AGS-1358386. This work is also supported by the Parker Solar Observatory contract with the Smithsonian Astrophysical Observatory through subcontract SV4-84017. We also acknowledge NSF PRAC award ACI-1144120 and related computer resources from the Blue Waters sustained-petascale computing project. Supercomputer allocations were also provided on SGI Pleiades by NASA High-End Computing Program award SMD-16-7570 and on Stampede2 by NSF XSEDE project MCA07S033. 

This work utilizes data from \textit{SOHO} which is a project of international cooperation between ESA and NASA. The HMI data have been used courtesy of NASA/\textit{SDO} and HMI science teams. The \textit{STEREO}/SECCHI data used here were produced by an international consortium of the Naval Research Laboratory (USA), Lockheed Martin Solar and Astrophysics Lab (USA), NASA Goddard Space Flight Center (USA), Rutherford Appleton Laboratory (UK), University of Birmingham (UK), Max-Planck-Institut for Solar System Research (Germany), Centre Spatiale de Li\`ege (Belgium), Institut d'Optique Th\'eorique et Appliqu\'ee (France), and Institut d'Astrophysique Spatiale (France). This work uses SOHO CME catalog which is generated and maintained at the CDAW Data Center by NASA and The Catholic University of America in cooperation with the Naval Research Laboratory.



\end{document}